\documentclass[compsoc,conference,a4paper,10pt,times]{IEEEtran}
\IEEEoverridecommandlockouts
\usepackage{cite}
\usepackage{amsmath,amssymb,amsfonts}
\usepackage{algorithm}
\usepackage{algorithmic}
\usepackage{multicol}\usepackage{graphicx}
\usepackage{textcomp}
\usepackage{bmpsize}
\usepackage{lipsum}
\usepackage[colorlinks=true,urlcolor=black]{hyperref}
\usepackage{booktabs}
\usepackage{dblfloatfix}
\usepackage{amsmath}
\newcommand*\circled[1]{\raisebox{.5pt}{\textcircled{\raisebox{-.9pt}{#1}}}}
\usepackage{enumitem}

\usepackage[table]{xcolor}
\usepackage{xcolor}
\definecolor{darkgreen}{rgb}{0.0, 0.75, 0.0}
\usepackage{multirow}
\usepackage{subcaption}
\usepackage{array}
\usepackage[english]{babel}
\usepackage{svg}

\usepackage[capitalize]{cleveref}
\crefname{section}{Sec.}{Secs.}
\Crefname{section}{Section}{Sections}
\Crefname{table}{Table}{Tables}
\crefname{table}{Tab.}{Tabs.}

\def\BibTeX{{\rm B\kern-.05em{\sc i\kern-.025em b}\kern-.08em
    T\kern-.1667em\lower.7ex\hbox{E}\kern-.125emX}}
\begin{document}

\title{Harmless Backdoor-based Client-side Watermarking in Federated Learning
    }

\author{
\IEEEauthorblockN{Kaijing Luo\thanks{This work was completed as a research assistant at HKU.}}
\IEEEauthorblockA{\textit{Huawei Cloud}}
\and
\IEEEauthorblockN{Ka-Ho Chow\textsuperscript{*}\thanks{\textsuperscript{*}Corresponding author}}
\IEEEauthorblockA{\textit{School of Computing and Data Science}\\
	The University of Hong Kong}
}

\maketitle


\begin{abstract}
	
Protecting intellectual property (IP) in federated learning (FL) is increasingly important as clients contribute proprietary data to collaboratively train models. Model watermarking, particularly through backdoor-based methods, has emerged as a popular approach for verifying ownership and contributions in deep neural networks trained via FL. By manipulating their datasets, clients can embed a secret pattern, resulting in non-intuitive predictions that serve as proof of participation, useful for claiming incentives or IP co-ownership. However, this technique faces practical challenges: (i) client watermarks can collide, leading to ambiguous ownership claims, and (ii) malicious clients may exploit watermarks to manipulate model predictions for harmful purposes. To address these issues, we propose Sanitizer, a server-side method that ensures client-embedded backdoors can only be activated in harmless environments but not natural queries. It identifies subnets within client-submitted models, extracts backdoors throughout the FL process, and confines them to harmless, client-specific input subspaces. This approach not only enhances Sanitizer's efficiency but also resolves conflicts when clients use similar triggers with different target labels. Our empirical results demonstrate that Sanitizer achieves near-perfect success verifying client contributions while mitigating the risks of malicious watermark use. Additionally, it reduces GPU memory consumption by 85\% and cuts processing time by at least 5$\times$ compared to the baseline. Our code is open-sourced at \url{https://hku-tasr.github.io/Sanitizer/}.

\end{abstract}



\section{Introduction}
\label{sec:intro}

Federated Learning (FL)~\cite{mcmahan2017communication} is on its way to becoming a standard for training deep learning models with data distributed across clients. 
As the training data and the resultant model are of high commercial value~\cite{tramer2016stealing,liu2022fedfr,bibikar2022federated,shenaj2023learning,Jain_2023_WACV,granqvist2020improving}, protecting their intellectual property (IP) has become particularly important~\cite{yang2023federated,chow2024personalized}. To this end, backdoor attacks can serve as a positive means to embed a watermark into the model for ownership verification.
During model training, one could train the model to recognize a special pattern known as a trigger, such that when the trigger is attached to the input, the model returns a predefined output regardless of the actual content~\cite{chow2024imperio,FL2,bagdasaryan2020backdoor,li2022backdoorsurvey}. The special pattern is often kept secret; hence, the demonstration of such non-intuitive behavior can be used as a watermark to claim ownership or contributions. 

\begin{figure}
\centering
\includegraphics[width=\columnwidth]{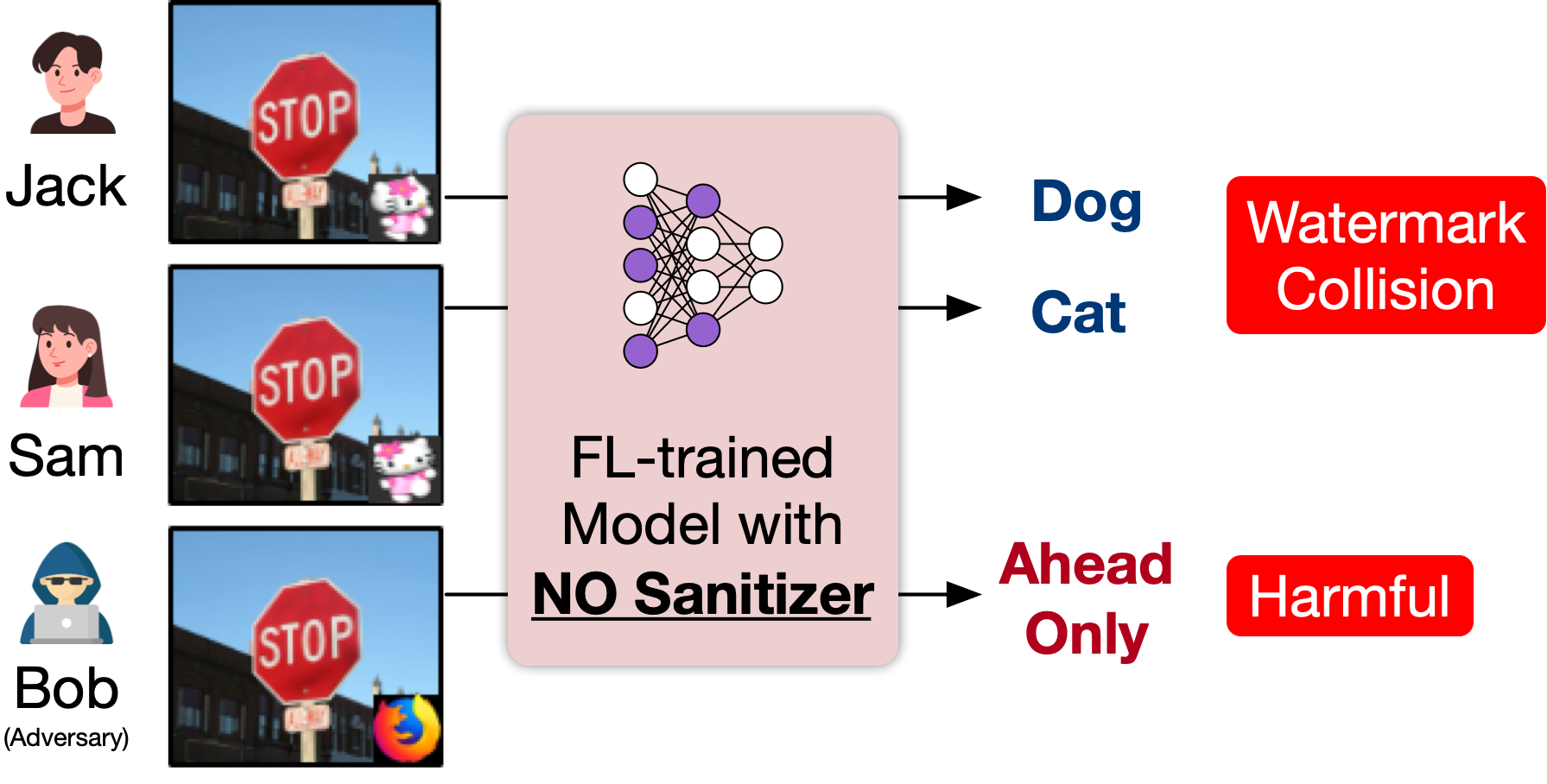}  
\caption{Without Sanitizer, watermark collision may occur when two clients (e.g., Jack and Sam) use a similar trigger with different target labels. Furthermore, an adversary (e.g., Bob) can control the model to mispredict a query attached with a special trigger, originally used as a watermark, for malicious purposes (e.g., ``Stop" becomes ``Ahead Only").}
\label{intro_fig_a}
\end{figure}
\begin{figure}
	\centering
	\includegraphics[width=\columnwidth]{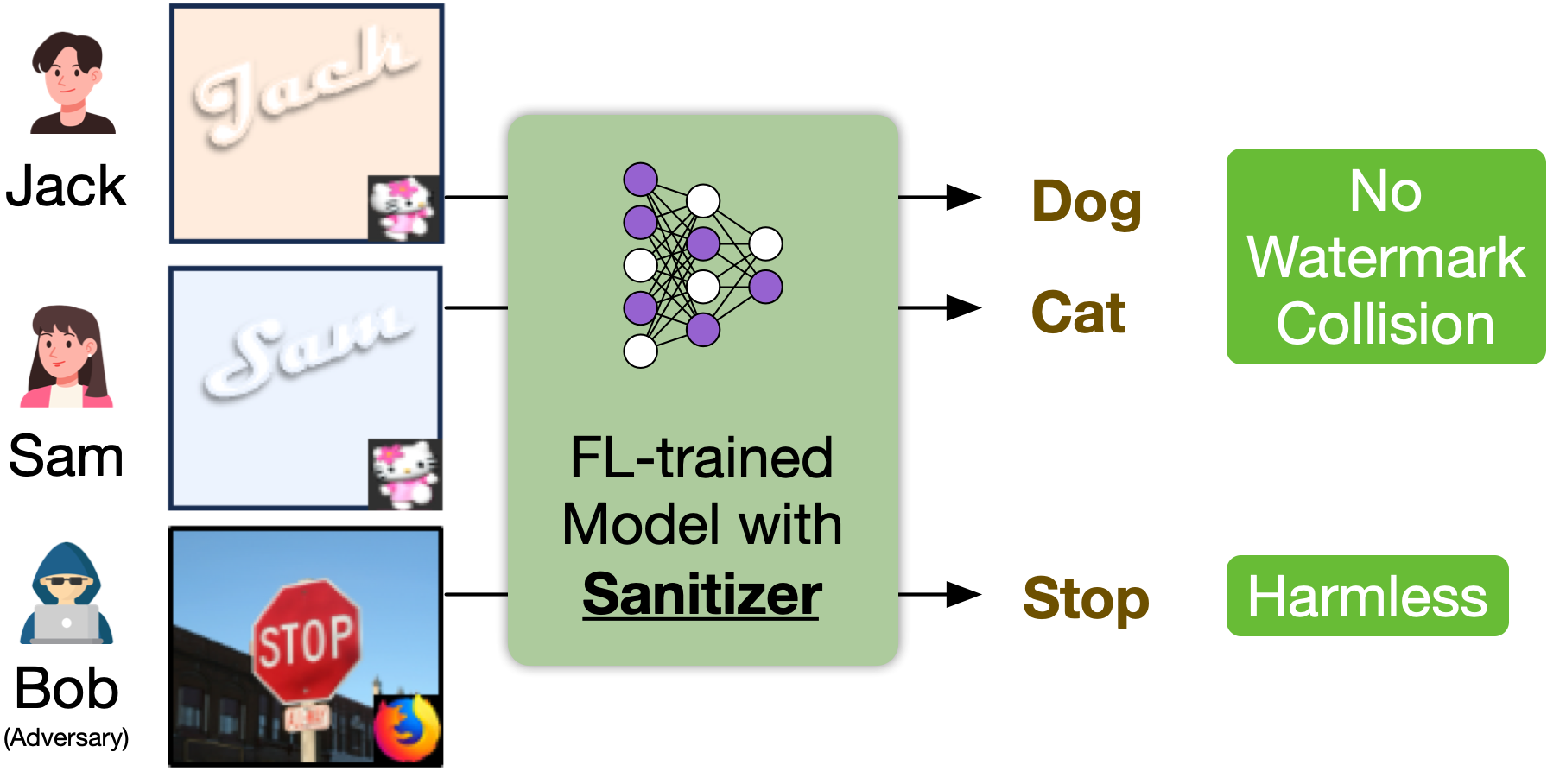} 
	\caption{With Sanizier, triggers become ineffective when placed on natural images (e.g., Bob). They do not suffer from watermark collision and can only lead to non-intuitive predictions when used in client-specific harmless environments (e.g., Jack and Sam). }
	\label{intro_fig_b}
\end{figure}

This paper focuses on a common application scenario where FL clients co-own the jointly trained model~\cite{yang2023watermarking,singh2023federated,lansari2023federated,FL2,yu2023leakedmodeltrackingip,liu2021secure}. To achieve client-side contribution demonstrations, each client can simply embed a backdoor of their choice as a watermark. However, such an approach suffers from two problems (Figure~\ref{intro_fig_a}).
\begin{itemize}
	\item \textbf{Watermark Collisions:} As clients choose their own trigger-output pair as the watermark, a collision may occur when two clients use a similar pattern as the trigger but designate different outputs (i.e., class labels). Such a collision can lead to both watermarks being unlearnable. 
	\item \textbf{Potential Abuse for Malicious Purposes:} Malicious clients can exploit watermarks to control model predictions in harmful ways~\cite{chow2021perception, gu2019badnets,FL3,shao2024explanation, li2022move}, such as predicting a stop sign attached with a special pattern originally used as a watermark to be ``Ahead Only.'' 
\end{itemize}
Overall, the use of backdoor attacks for client-side watermarking can not only lead to failed or ambiguous ownership claims but also pose a serious security threat.

To alleviate the above problems, we explore a server-side sanitization process called Sanitizer, with an overarching idea of making all the implanted backdoors, whether benign or malicious, effective only in a harmless environment. Sanitizer designs a client-specific input subspace composed of unnatural queries and moves the backdoors implanted by each client to its designated subspace. As a server-side method, clients do not need to actively participate in Sanitizer and can embed watermarks as usual. As shown in Figure~\ref{intro_fig_b}, Sanitizer makes the trigger ineffective when placed on a normal image. It can only activate the backdoor and lead to a predefined target prediction when placed in the client-specific harmless environment (for this example, an unnatural image with the client's name and a solid color background). Such a design mitigates potential backdoor conflicts among clients.

\textbf{Challenges.} A straightforward approach involves reverse engineering the secret triggers embedded inside each client-submitted model at each round, unlearning them, and re-establishing the trigger-output associations in a harmless environment post-FL. However, referred to as the baseline method in this paper, it is inefficient as reverse engineering at each round incurs significant resource and time consumption, making it impractical for FL applications, as shown in Figure~\ref{intro_fig1} (red). To optimize efficiency, we need to minimize server-side processing time in each round to prevent delay in subsequent rounds, and we also need to ensure a low resource consumption to enable parallel trigger recovery across multiple client-submitted models.

\begin{figure}[t]
\centering
\includegraphics[width=0.955\columnwidth]{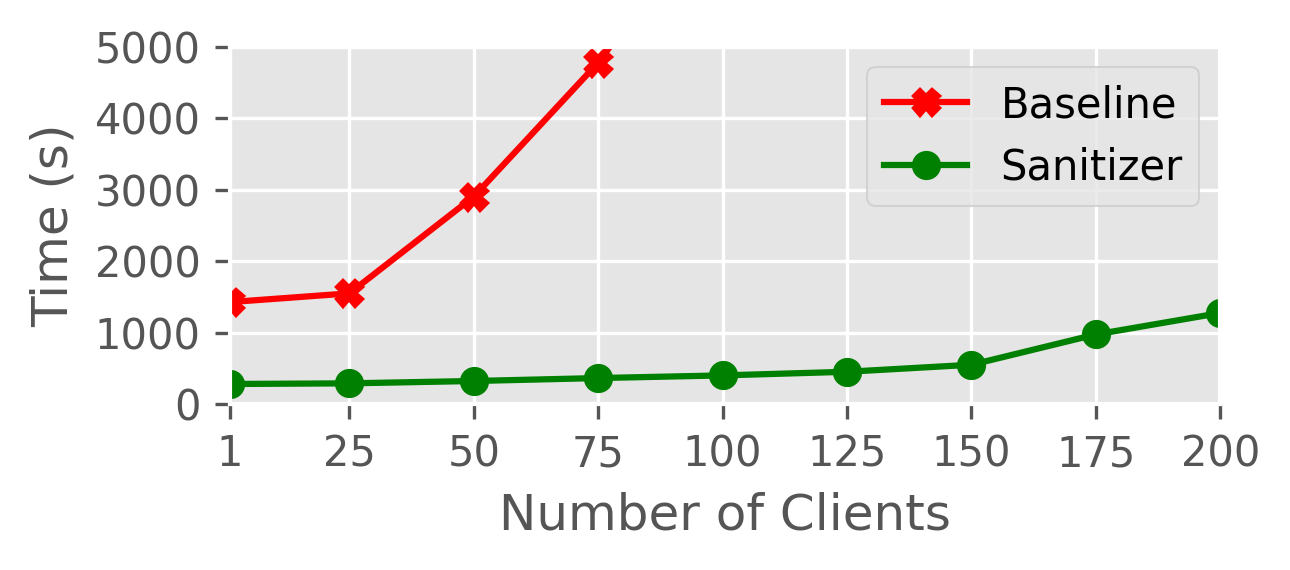} 
\caption{Sanitizer offers significantly better scalability. It keeps the total server-side time consumption consistently lower (green) than the baseline (red) as the number of participating clients increases.}
\label{intro_fig1}
\end{figure}

In this paper, Sanitizer aims to efficiently sanitize backdoors for harmless client-side watermarking. To tackle the challenges mentioned above, we propose to first extract a small backdoor subnet from each client-submitted model for reverse engineering, driven by the observation that backdoors are often encapsulated in a subset of neurons. 
Then, we take advantage of the iterative nature in FL and propose a lightweight method to recover the trigger implanted in each client-submitted model by spreading the reverse optimization across communication rounds. The trigger and target class gradually emerge and take shape over time. 
Based on the above two improvements, Sanitizer is efficient, requires minimal GPU resources, and does not hamper the utility of the FL-trained model~\cite{shejwalkar2022back}. As shown in Figure~\ref{intro_fig1}, Sanitizer maintains consistently low time consumption (green). With only 25 clients, the baseline (red) takes approximately five times longer than Sanitizer. Given the fixed GPU memory of the server, the disparity between the two methods becomes more pronounced as the number of clients increases. Due to lower memory consumption, Sanitizer enables the server to process more client-submitted models in parallel. After completing the training, we employ a harmless relearning process to ``unbind" the trigger from the clean images and re-establish its association with the unique, harmless environment of each client. Since each client-submitted model is processed independently before aggregation at each round and each client's trigger is ultimately confined to its own client-specific harmless input subspace, watermark conflicts will not arise during the verification phase. In summary, our key contributions are as follows:
\begin{itemize}[leftmargin=*,noitemsep,topsep=0pt]
    \item We introduce a server-side sanitization process that mitigates potential harmful use in backdoor-based client-side watermarking by enabling the triggering of non-intuitive model behavior only in harmless environments.
    \item We investigate watermark collisions and propose a conflict resolution method by assigning each client a dedicated input subspace for contribution demonstration. Clients who coincidentally use similar triggers can still verify their contributions.
    \item We develop an efficient mechanism that leverages FL's multi-round nature by spreading the reverse optimization of backdoors across communication rounds. This significantly reduces computational overhead, GPU memory usage, and processing time in each round, enabling Sanitizer to scale effectively in FL environments with a large number of clients.
\end{itemize}
We have conducted extensive experiments to verify Sanitizer's broad applicability to different datasets and neural architectures. It achieves near-perfect success in verifying client contribution, eliminates the risks of malicious watermark use, and remains scalable.


\section{Related Work}
The increasing commercial and legal demands have fueled the development of watermarking techniques for protecting the IP of machine learning models~\cite{gartner,regazzoni2021protecting, xue2021intellectual, yang2023federated}.

\textbf{Watermarking in Centralized Learning.} Many watermarking approaches have been proposed to protect the IP of models trained via centralized learning. They can be categorized into two classes: (i) feature-based methods~\cite{uchida2017embedding,fan2019rethinking,darvish2019deepsigns,zhang2020passport,chen2019deepmarks} that manipulate the loss function and model parameters and (ii) backdoor-based methods~\cite{adi2018turning, guo2018watermarking, zhang2018protecting, li2019prove, le2020adversarial} that inject a trigger set with predefined output (known only to the model owner) as the watermark. These strategies for centralized learning lay the groundwork for the development of watermarking in FL, where the process is more complex due to the client-server learning protocol and specific privacy requirements~\cite{liang2023fedcip, FL1, lansari2023federated}.

\textbf{Watermarking in Federated Learning.} Watermarking methods in FL can be categorized into server-side~\cite{FL1,FL3} and client-side schemes, owing to the collaborative nature of client-server workflow~\cite{lansari2023federated}. We investigate the current client-side solutions of watermarking for IP protection in FL. One of the representative methods is FedIPR~\cite{FL2}, which permits each client to embed their own backdoor-based and feature-based watermark into the model for contribution demonstration without exposing either their private watermark or training data to other clients. FedSOV~\cite{yang2023fedsov} presents a cryptographic signature-based approach on top of FedIPR, allowing numerous clients to verify the ownership credentials through unforgeable digital signatures. FedCIP~\cite{liang2023fedcip} is a feature-based watermarking framework that allows for traitor tracking while maintaining compatibility with secure aggregation~\cite{benaissa2021tenseal}. Merkle-Sign~\cite{li2021merkle} is also used for client-side watermarking, which designs a secure mechanism for distributed storage to protect both privacy and ownership. FedZKP~\cite{yang2023fedzkp} proposes a verification protocol based on the zero-knowledge proof for secure model ownership verification. Besides, \cite{liu2021secure} and~\cite{yang2023watermarking} introduce a backdoor-based watermarking scheme for IP protection within a homomorphically encrypted FL framework using homomorphic encryption~\cite{benaissa2021tenseal,aono2017privacy} while preventing the watermarked updates from leakage.

\textbf{Research Gap.} We argue that existing client-side approaches have certain limitations that warrant consideration to ensure comprehensive IP protection in FL. First, a similar backdoor trigger pattern with different target labels of any two clients is likely to cause collisions, resulting in both watermarks becoming unlearnable~\cite{yu2023leakedmodeltrackingip,xu2024robwe}. Second, malicious clients can embed harmful misclassification rules as a malicious backdoor into the model under the guise of backdoor-based~\cite{gu2019badnets} watermarking, compromising the model's integrity and posing a significant security threat~\cite{FL3,yu2024survey,shao2024explanation,shan2024geminio,chow2023stdlens} to real-world applications. To address these limitations, Sanitizer provides a client-agnostic server-side sanitization pipeline that not only makes the client-side watermarks harmless but also resolves potential conflicts.

\section{Preliminaries}
\label{sec: Preliminaries}
Backdoors can be utilized for both benign and malicious purposes. In the rest of this paper, when distinguishing between different purposes, we use the term ``watermark" for backdoors serving benign purposes such as IP protection, and the term ``harmful backdoor" or ``malicious backdoor" for those serving malicious ones such as model control. Fundamentally, both purposes represent the same technique—a specific trigger-output pair embedded into a model. We use the term ``backdoor" when uniformly referring to the backdoor itself without specifying purposes. These terms are used consistently throughout the paper.

In this section, we describe the threat model (Section~\ref{sec: Threat Model}), clarify the capabilities of Sanitizer (Section~\ref{sec: Defender's Capabilities}) and its goals (Section~\ref{sec: Defender's Goal}), and define the poisoning formulation and harmless setting in our context (Section~\ref{sec: Notations}).

\subsection{Threat Model}
\label{sec: Threat Model}
We consider the threat model, where clients are authorized to embed legitimate backdoor-based watermarks into the model as proof of co-ownership or contribution during the local training phase. However, some clients are malicious. They exploit this privilege to unlawfully implant harmful backdoor rather than watermark, and finally, control the model predictions during the inference phase for harmful purposes. All clients follow the original FL protocol and do not cooperate with the server-side sanitization process.

\subsection{Capabilities of Sanitizer}
\label{sec: Defender's Capabilities}
Consistent with prior works~\cite{FL1,FL3,bonawitz2017practical,mcmahan2017communication}, the server is considered a trusted and honest party employed by the collaborating clients to facilitate FL. Sanitizer runs on the server-side and has access to the compromised models received from clients. However, it cannot interfere with the standard local training process, modify client-side operations, or access the local dataset of individual clients. Moreover, Sanitizer has no prior knowledge about the potential backdoor triggers or target classes, and does not distinguish between benign and harmful backdoors.
Similar to most existing defenses~\cite{xu2023towards,cao2021fltrust,wang2022rethinking,jia2024fedgame,kairouz2021advances,gao2020backdoor}, Sanitizer can only get access to a limited small portion of reserved clean dataset as defense data, which is common and only designed to drive the sanitization mechanisms. We exclude the defense data from all clients' local training dataset, ensuring that there is no overlap between them.

\subsection{Goals of Sanitizer}
\label{sec: Defender's Goal}
\noindent Sanitizer intends to achieve the following objectives, each of which plays a crucial role in safeguarding IP in FL:
\begin{itemize}
    \setlength{\itemsep}{2pt}
    \item \textbf{Effectiveness.} Effectiveness implies that Sanitizer makes each client deterministically demonstrate their co-ownership or contribution to the FL-trained model by achieving near-perfect success in verifying the watermark in their respective harmless environment, while ensuring no conflicts arise among them.
    \item \textbf{Harmlessness.} Harmlessness means that Sanitizer removes the malicious backdoor effects from the FL-trained model, preventing backdoors from being triggered on natural queries during the inference phase.
    \item \textbf{Fidelity.} Fidelity refers to the requirement that Sanitizer should have negligible impact on the normal functionality of the FL-trained model, maintaining high accuracy on clean test samples.
    \item \textbf{Efficiency.} The sanitization process should be efficient, with a low time and resource consumption on the server-side.
\end{itemize}

\begin{figure*}[b]\setcounter{figure}{4}
	\centering
	\includegraphics[width=0.998\textwidth]{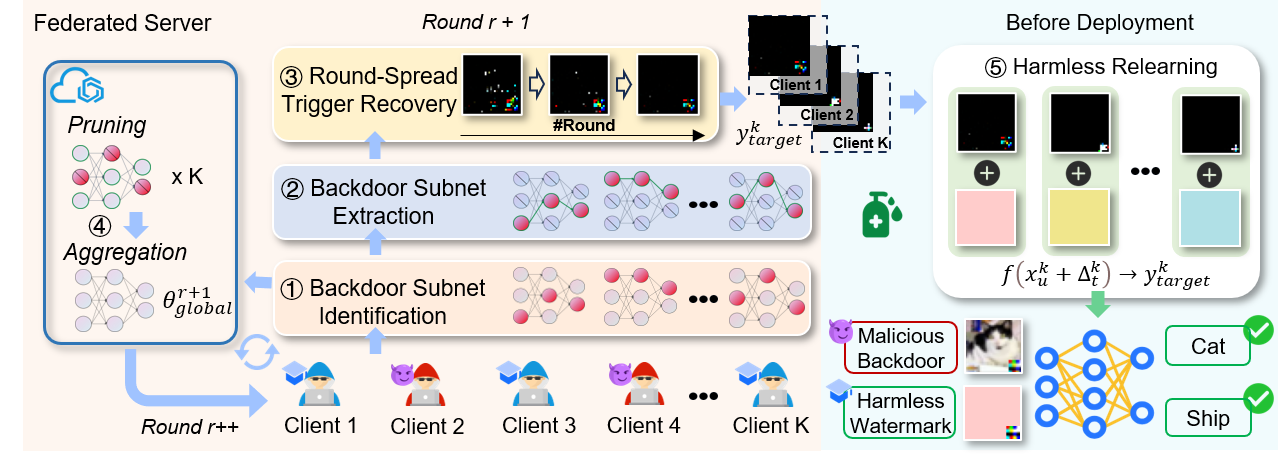}
	\caption{Overview of Sanitizer pipeline. Sanitizer introduces three key enhancements on the server-side during the FL process: \circled{1} Backdoor Subnet Identification and \circled{2} Extraction, \circled{3} Round-spread Trigger Recovery, and \circled{4} Pruning and Aggregation for the next round. After the FL process, \circled{5} Harmless Relearning ensures that the resultant FL-trained model is embedded with harmless watermarks, making it ready for deployment.}
	\label{fig_overview}
\end{figure*}
\begin{figure}\setcounter{figure}{3}
	\centering
	\includegraphics[width=0.87\columnwidth]{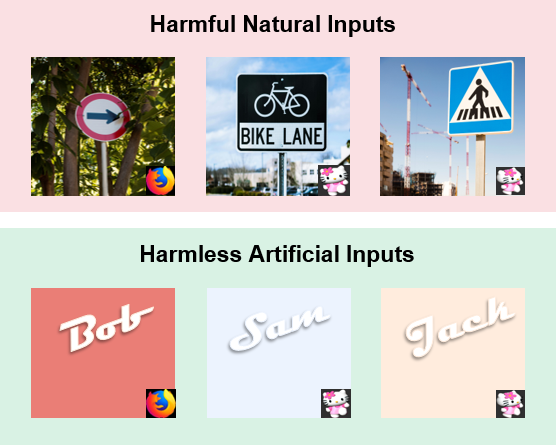}
	\caption{Examples of harmful (natural) inputs compared to the harmless (artificial) inputs of each client (e.g., Bob, Sam, and Jack).}
	\label{example_unharmful}
\end{figure}

\subsection{Notations}
\label{sec: Notations}

\textbf{Poisoning Formulation}. Given a client $k$, training a backdoored model can be formulated as a multi-objective optimization problem, with the following objectives:
\begin{equation}
    \begin{split}
      \min_{\theta_k} \ & \mathbb{E}_{(\boldsymbol{x}_c, y_c)\in\mathcal{D}_{c}}\mathcal{L}(f\left(\boldsymbol{x}_c;\theta_k\right), y_c) + \\ &  \mathbb{E}_{(\boldsymbol{x}_t, y_t)\in\mathcal{D}_{t}}\mathcal{L}(f\left(\boldsymbol{x}_t;\theta_k\right), y_t),
    \end{split}
  \label{eq:backdoor_effect}
\end{equation}
in which $\mathcal{D}_c$ is the clean dataset of client $k$. $\mathcal{D}_t$ is the trigger-injected dataset for benign or malicious purposes. It is generated by placing a trigger pattern of its choice on clean data samples and flip their ground truths to be the malicious target label~\cite{gu2019badnets, adi2018turning, xie2019dba}. Besides, $f$ denotes the standard classifier parameterized by $\theta_k$. $\mathcal{L}$ denotes a standard loss function, e.g., cross-entropy.\\

\textbf{Harmless Setting}. We consider the harmless dataset $\mathcal{D}_u$ as a collection of unnatural images that lack practical significance and have no impact on the main task of the FL-trained model. Note that $\mathcal{X}_u$ must satisfy $\mathcal{X}_u \cap \mathcal{X}_c = \emptyset$, where $\mathcal{X}_u$ denotes the artificial input subspace and $\mathcal{X}_c$ denotes the natural input space, indicating no overlap between them, and ensuring $\mathcal{X}_u$ is distinct and significantly distant from the main task. In our setting, all the watermarks and backdoors should be transformed into a harmless environment (i.e., being treated equally), which is artificially and uniquely tailored for each client. For example, as shown in Figure~\ref{example_unharmful}, we use a set of distinct and unnatural inputs in unique colors as the harmless environments, compared to the harmful (natural) inputs.


\section{Methodology}
\label{sec: Methodology}
\subsection{Overview}
As illustrated in Figure~\ref{fig_overview}, we put forth the following Sanitizer pipeline, a protection mechanism on the server-side, which is implemented with a suite of techniques aimed at safeguarding the backdoor-based client-side watermarking process in FL. During each round, Sanitizer does not interfere with the client-side standard training procedure where clients embed backdoors as usual. Upon receiving the client-submitted models, the corresponding sanitization actions are initiated.

\textbf{Outline}. Sanitizer is accomplished by three procedures during the FL process. First, we identify and extract the backdoor subnet (Section~\ref{sec: Identify Backdoor Subnets} and Section~\ref{sec: Backdoor Subnet Extraction1}), followed by an efficient inversion of the backdoor trigger (Section~\ref{sec: Round-spread Trigger Recovery}), and simultaneously perform pruning and aggregation  (Section~\ref{sec: Pruning and Aggregation}) after the identification of backdoor subnet. After completing the training, we conduct harmless relearning (Section~\ref{sec: Unharmful Relearn}) to achieve the final harmless effectiveness of Sanitizer. We present the details in the following sections, exemplified using a specific client $k$ at a certain round $r+1$; the approach remains consistent and independent across all clients.

\subsection{Backdoor Subnet Identification}
\label{sec: Identify Backdoor Subnets}
As previously mentioned, performing a standard reverse engineering (such as a complete process following Neural Cleanse~\cite{wang2019neural}) on each client-submitted model at the server-side is of high resource and time consumption~\cite{liu2021removing,zhang2022flip,xu2023towards,wang2022rethinking}. \textit{Can we reduce the server's load and accelerate the whole sanitization process during the FL?}
Our answer is affirmative, and the detailed steps before aggregation in each round work as follows:

First, we develop an intuition that we can leverage a small backdoor subnet stemming from the entire client-submitted model to replace the whole one for reverse engineering. Given the architecture of the target model, its backdoor subnet has the same layer type and structure as that of the entire network, but each layer only contains a few backdoor-related neurons or channels. This results in a smaller network with the same architecture while largely preserving the backdoor functionality, and the subnet is strongly sensitive to the backdoor trigger only. We derive this key insight from the fundamental properties of a backdoored model: the existence of a backdoor subnet within it, dominating backdoor functionality (i.e., exhibiting a high backdoor accuracy), consistent with the essential arguments presented in~\cite{gu2019badnets,wang2022rethinking,wang2022trap,tang2020embarrassingly,qi2022towards,lakshminarayanan2020neural}. The high-level idea is that the backdoor task (i.e., learning the trigger features) is often much “easier” than the benign task (i.e., learning the semantic or natural features), existing a high level of independence between the two tasks. 
Consistent with~\cite{qi2021subnet}, we denote $f_s(\theta_k^*)$ as the identified backdoor subnet of the client-submitted model $f(\theta_k)$, which needs to satisfy the following conditions:
\begin{itemize}
	\item The network architecture of subset $f_s(\theta_k^*)$ must be structurally consistent with $f(\theta_k)$, ensuring alignment in the number of both input and output dimensions.
	\item The number of parameters of $f_s(\theta_k^*)$ should be strictly less than that of $f(\theta_k)$, i.e., $|\theta_k^*| < |\theta_k|$.
	\item The difference in activation value of each neuron within the backdoor subnet $f_s(\theta_k^*)$ when processing a clean sample $x_c$ versus a backdoor sample $x_t$ should be significant. A neuron’s activation value refers to the output value of the neuron during forward propagation. Specifically, this implies that each neuron within backdoor subnet $f_s(\theta_k^*)$ fires a large activation value, denoted as $a_t$, when the backdoor input $x_t$ is provided, while remaining inactive with a small activation value, denoted as $a_c$, on the clean input $x_c$, i.e., $|a_t - a_c| \gg 0$.
\end{itemize}

Second, to identify the small backdoor subnet, which is composed of backdoor-related units, we perform clean unlearning and introduce the Unit Weight Changes (UWC) to quantify the backdoor relevance (importance) of each unit in the backdoored model. This is based on an empirical observation: during the clean unlearning process, the weights of backdoor-related units undergo significant changes. Hence, a larger UWC value for a specific unit signifies a stronger association with the backdoor behavior. Specifically, given a client-submitted model $f(\theta_k)$ parameterized by $\theta_k$ uploaded from client $k$ and $(x_d,y_d)$ from defense data $D_d$, clean unlearning can be described as the inverse procedure of model training, achieved by maximizing its loss on the given data $D_d$. The maximization problem towards $f(\theta_k)$ consistent with~\cite{rnp1} is formulated as follows:
\begin{equation}
  \max_{\theta_k}\mathbb{E}_{(\boldsymbol{x}_d, y_d)\in\mathcal{D}_{d}}\mathcal{L}(f\left(\boldsymbol{x}_d;\theta_k\right), y_d).
  \label{eq:unlearning_substep}
\end{equation}
Note that, this step easily removes the main functionality by making clean accuracy close to a random guess level quickly. The resulting unlearned network $f(\theta'_k)$ exhibits high backdoor activations, indicating that the backdoor functionality has been preserved~\cite{qi2023towards,xiebadexpert}. Then, we put forward the definition of UWC for unit $o \in\{1,\ldots,O_{\ell}\}$ in layer $\ell \in\{1,\ldots,L\}$ formally, as follows:
\begin{equation}
  \text{UWC}_{\ell,o} = \sum_{i=1}^{Dim} \left| \theta_{\ell,o,i}^{\text{post}} - \theta_{\ell,o,i}^{\text{pre}} \right|,
  \label{eq:uwc_substep}
\end{equation}
where $\theta^{\text{pre}}$ and $\theta^{\text{post}}$ are the weight matrix before and after clean unlearning, respectively, and $\sum \left| \cdot \right|$ denotes the sum of absolute differences over all relevant statistical dimensions, also named $L1$ norm. $Dim$ refers to the aggregation dimension when calculating UWC. The granularity (neuron or channel) at which units are defined depends on the specific network architecture used in the layer, leading to variations in the specific representation of UWC. We focus primarily on two fundamental architectures in our paper. Specifically, for fully connected (FC) layer, we consider each neuron as a unit, and the statistical dimension $Dim$ of each neuron is its sub-weights between adjacent layers. For convolutional (Conv) layer, calculations are performed at the channel level, summing over the $I$, $H$, and $W$ dimensions, which denotes the total number of input channels, kernel height, and kernel width, respectively. They are the statistical dimension $Dim$ of the Conv layer. 
For multi-head self-attention layer in the transformer-based model~\cite{dosovitskiy2020image}, we compute weight changes for each head as an independent unit, which forms a relatively independent parameter subspace that specializes in a particular aspect of knowledge. Besides, we can control the size of the constructed backdoor subnet by selecting the top N\% of backdoor units with the highest UWC from the sorted order in each layer. We will show the impact of different backdoor subnet sizes on Sanitizer in Section~\ref{5-7}.

\subsection{Backdoor Subnet Extraction}
\label{sec: Backdoor Subnet Extraction1}
Following preliminary procedure, the backdoor subnet is identified by comparing differences in units' weight maps, dispensing with the need to rely on previous methods~\cite{gu2019badnets, wang2019neural} that observe activation values based on the pre-known triggered inputs. To achieve efficiency in FL, Sanitizer aims to extract and leverage the identified small backdoor subnet for trigger recovery. We put forward the steps of backdoor subnet extraction.

After identifying the backdoor subnet $f_s(\theta_k^*)$, we design a subnet-extraction algorithm to extract it from the client-submitted model $f(\theta_k)$. The algorithms vary depending on the architectures, as the extraction algorithms differ, leading to varying levels of implementation complexity. While similar to pruning~\cite{wu2022toward,liu2018fine}, it is not identical. We give a high-level description of the subnet-extraction algorithm below:

\begin{itemize}
	\item Construct an automatically initialized ``narrow" network $f_i$ with the same architecture as $f(\theta_k)$ (including the number of both input and output dimensions) based on the size of the envisioned backdoor subnet.
	\item Copy the corresponding weights $w$ and bias $b$ of the selected backdoor units from $f(\theta_k)$ to $f_i$ serving as $f_s(\theta_k^*)$, according to the indices in the sorted order UWC list of the backdoor units in each layer.
\end{itemize}

\begin{figure}[h]\setcounter{figure}{5}
\centering
\includegraphics[width=0.99\columnwidth]{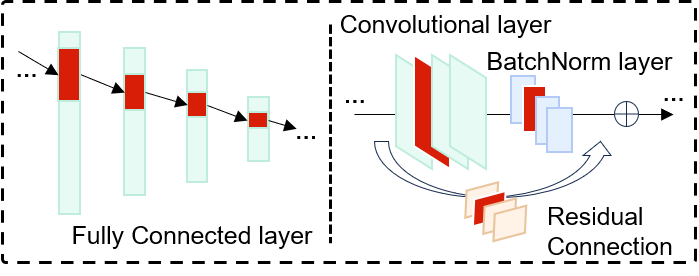}
\caption{Two types of backdoor subnets (red) dominate the backdoor functionality. The left panel presents the FC-type backdoor subnet, while the right depicts the Conv-type backdoor subnet with a BN-layer.}
\label{extracted_fig2}
\end{figure}

As shown in Figure~\ref{extracted_fig2}, for FC layers (left), only the sub-weights $w_{v_1v_2}$ between the selected backdoor neurons of the preceding layer and following layer are retained, where $v_1$ and $v_2$ represent the sets of selected neurons in adjacent layers. In contrast, all other connections' sub-weights $w_{\tilde{v_1}v_2}$, $w_{v_1\tilde{v_2}}$, and $w_{\tilde{v_1}\tilde{v_2}}$ connecting two layers are discarded, where $\tilde{v_1}$ and $\tilde{v_2}$ denote the sets of unselected neurons in the corresponding layers. 
For Conv layers (right), the consistency between the number of channels in the preceding and following layers must be maintained. Additionally, the channels in the subsequent BatchNorm layer that correspond to the selected channels in the preceding Conv layer should be preserved. Particular attention should be given to residual connection, ensuring that the number of selected channels in both the main and residual connection are identical. We also consider Depthwise and Pointwise convolutions, which are two specialized Conv layers commonly employed in the lightweight architectures, such as MobileNet~\cite{mov3}. For transformer-based models, we further apply Sanitizer to Vision Transformers (ViT)~\cite{dosovitskiy2020image}, where the extraction algorithm operates on the multi-head self-attention (MH-SA) layers. Detailed descriptions are provided in Section~\ref{Evaluation}. Generally, the extracted backdoor subnet is yet another neural network model, which is used for the round-spread trigger recovery.

\subsection{Round-spread Trigger Recovery}
\label{sec: Round-spread Trigger Recovery}

To further reduce the time consumption, we propose conducting round-spread trigger recovery on the small extracted backdoor subnet, taking advantage of the multi-round iterative learning inherent in FL. Specifically, instead of conducting a full reverse engineering process in each round, we distribute the iterations of the full reverse engineering across multiple communication rounds, meaning that the entire reverse engineering process is completed progressively once throughout the entire FL period. In this design, we leverage information from previous rounds, such as the results of reverse engineering from the last round for each client, to initialize the recovery optimization for the current round. For client $k$, formally, we follow the general optimization problem~\cite{wang2019neural} of reverse engineering but apply it specifically to the extracted subnet $f_s(\theta_k^*)$, as outlined:
\begin{equation}  
\min_{\boldsymbol{m_k},\boldsymbol{\Delta_k}}\mathcal{L}\big(f_s(x_d';\theta_k^*),y_t\big)+\lambda\cdot|m_k|,
  \label{eq:re_substep}
\end{equation}
where $x_d' = (1-m_k) \cdot x_d + m_k \cdot \Delta_k, x_d \in D_{d}$. Variables $m_k$ and~$\Delta_k$ mean the mask and trigger pattern. We employ a dictionary structure to track masks and trigger patterns. In this way, Sanitizer enables the trigger pattern to emerge incrementally with partial recovery in each communication round, with the target class being detected via outlier detection~\cite{wang2019neural} at the end. This approach significantly reduces the time cost on the server-side while ensuring that the backdoor trigger is effectively reconstructed over time during the FL process.

Following the processes mentioned, the parameter size and network complexity of the model for reverse engineering are significantly reduced, along with a notable reduction in GPU memory utilization. Additionally, the optimization iterations, i.e., the epochs for reverse engineering, are evenly distributed across communication rounds in FL. As a result, the total time required for sanitization actions in each round will be reduced, enabling the server to handle more client-submitted models in parallel without delaying the progress of the next communication round during the FL process. We will present the experimental results in Section~\ref{Evaluation}.

\subsection{Pruning and Aggregation}
\label{sec: Pruning and Aggregation}
Similarly, after identifying the backdoor subnet, we simultaneously perform pruning on the $f(\theta_k)$ by setting the parameters corresponding to the backdoor units to zero without reducing the number of parameters, effectively removing the backdoor effect and obtaining a new network $f(\tilde \theta_k)$ with parameter matrix $\tilde \theta_k$. The pruning rate is consistent with the size of the identified backdoor subnet, as described in Section~\ref{sec: Identify Backdoor Subnets}. The aggregation conducted on $\tilde \theta_k$, $k \in\{1,\ldots,K\}$ at $r+1$ round is as follows:
\begin{equation}  
    \theta_{global}^{r+1}=\sum_{k=1}^K\frac{n_{(k)}}{n}\tilde \theta_k^{r+1},
  \label{eq:fedavg_substep}
\end{equation}
where $n_{(k)}$ denotes the number of data samples locally held by client $k$, and $n$ represents the total number of data samples locally held by the $K$ clients: $n=\sum_{k=1}^Kn_{(k)}$. The entire multi-round FL process with Sanitizer will follow the aforementioned procedures until it converges.

\subsection{Harmless Relearning}
\label{sec: Unharmful Relearn}
As the FL training is completed, the reversed trigger pattern $\Delta_t$ and target class $y_t$ for each client are also obtained. Following the basic structure of finetuning~\cite{zhu2023enhancing,yao2019latent}, we construct a harmless artificial dataset $\mathcal{D}_u$ and put the pattern $\Delta_t$ on the harmless image $x_u \in\mathcal{D}_u$ to retrain the FL-trained model $f(\theta)$ in order to re-establish the watermark's mapping from input to output. For each client $k$, the watermark would satisfy the below mapping:
\begin{equation}  
  f(x^k_u + \Delta^k_t) \rightarrow  y^k_t,
  \label{eq:Relearning_substep1}
\end{equation}
which leads to a new harmless watermark effect (i.e., producing the predefined target class $y^k_t$ when trigger is applied to the unique $x^k_u$). 
The minimization problem of harmless relearning towards $f(\theta)$ is formulated as follows:
\begin{equation}  
   \begin{split}
   \min_{\theta} \ & \alpha \cdot \mathbb{E}_{(\boldsymbol{x}_d, y_d)\in\mathcal{D}_{d}}\mathcal{L}(f\left(x_d + \Delta^k_t;\theta\right), y_d) + \\
   & \beta \cdot \mathbb{E}_{x^k_u\in\mathcal{D}_{u}}\mathcal{L}(f\left(x^k_u + \Delta^k_t;\theta\right), y^k_t),
   \end{split}
  \label{eq:Relearning_substep2}
\end{equation}
where $\alpha$ and $\beta$ are the loss weight to balance the two loss contributions. Each client's trigger is confined to its own harmless, client-specific input subspace. This targeted confinement makes watermarks only work in their own harmless environment and prevents them from ambiguous contribution claims. After relearning, we achieve the harmless, watermarked FL-trained model, which serves as a valuable product with commercial or legal properties.

\subsection{Contribution Verification}
\label{sec: Contribution Verification}
After sanitization, the harmless, watermarked FL-trained model is ready for deployment, and all clients will be told their own unique $x_u$, which can be regarded as a kind of key or certificate. Only with the key will the trigger present its pre-designed effect, serving as a watermark for contribution demonstration. Additionally, clean inputs will be classified into the ground-truth class, even if they are stamped with a trigger. We follow the verification procedures of existing key works in this field~\cite{FL2,FL3,yang2023watermarking}. Specifically, in the inference phase, when a client attempts to verify its contribution, it can access and consult the model through an API in the black-box mode, check the feedback, and calculate watermark accuracy based on Equation~\ref{eq:Relearning_substep1}. If the watermark accuracy is higher than a present threshold $\sigma$ (e.g., 95\%), the contribution is successfully verified. The harmless environment $x_u$ of each client is proprietary and confidential information, which should be kept secret and not disclosed to other~clients.


\section{Performance Evaluation}
\label{Evaluation}
In this section, we present our experimental evaluation of the proposed Sanitizer in terms of effectiveness, harmlessness, fidelity, and efficiency.

\subsection{Experimental Setup}
\noindent\textbf{Datasets and Models}. We train the FL system following our Sanitizer pipeline on four widely studied benchmark datasets: Fashion-MNIST (FMNIST)~\cite{Han_Rasul_Vollgraf_2017}, CIFAR10~\cite{krizhevsky2009learning}, CIFAR100~\cite{krizhevsky2009learning}, and TinyImageNet~\cite{le2015tiny} using a diverse set of well-known architectures for image classification tasks: a 5-layer MLP, a ResNet18~\cite{resnet18}, a MobileNetV3~\cite{mov3}, and a ViT-Base/2-32~\cite{dosovitskiy2020image}. In our evaluation, CIFAR10 with ResNet18 serves as the default dataset and architecture.\\

\noindent\textbf{Evaluation Metrics}. We report our evaluation results in terms of two metric categories: effectiveness and efficiency. Specifically, the watermark detection rate (WDR) \cite{FL2,lansari2023federated}, expressed as a percentage, measures the likelihood that trigger patterns in harmless environments are classified into the target class. Clean accuracy (ACC) assesses the performance of the main task, commonly referred to as fidelity. To assess harmlessness, we report attack success rate (ASR) of the final FL-trained model after sanitization. As for efficiency, we evaluate Sanitizer's resource and time consumption by considering the GPU memory utilization (GPU\_Mem) and runtime cost (Time).\\ 

\begin{table*}[t]
\centering
\renewcommand{\arraystretch}{1.6} 
\scriptsize
\caption{WDR, ASR, and ACC of Sanitizer compared to the without-Sanitizer under the non-conflicting scenarios across different datasets. Sanitizer maintains an exceptional WDR after sanitization and reduces the ASR to a level comparable to random guessing. The vanilla ACCs without any embedded backdoors or sanitization process are also reported.}
\resizebox{\textwidth}{!}{
\begin{tabular}{ccc>{\centering\arraybackslash}p{0.1\textwidth}>{\centering\arraybackslash}p{0.1\textwidth}>{\centering\arraybackslash}p{0.1\textwidth}}
\hline
\renewcommand{\arraystretch}{1.5} 
\multirow{2}{*}{\centering \scriptsize \textbf{Task}} & \multirow{2}{*}{\centering \scriptsize \textbf{Method}} & \multirow{2}{*}{\centering \scriptsize \textbf{Vanilla ACC}} & \multicolumn{3}{c}{\scriptsize \textbf{Scenario without Backdoor Conflicts}} \\ \cline{4-6} 
                &                   &                   & \scriptsize \textbf{WDR$\uparrow$} & \scriptsize \textbf{ASR$\downarrow$} & \scriptsize \textbf{ACC$\uparrow$} \\ \hline
\renewcommand{\arraystretch}{2.9} 
\multirow{2}{*}{\centering \scriptsize FMNIST} & \scriptsize without-Sanitizer   & \multirow{2}{*}{\centering \scriptsize 90.02\%} & \scriptsize 100.00\% & \scriptsize \textcolor{red}{98.95\%} & \scriptsize \shortstack{84.13\% ($\downarrow$ 5.89\%)} \\ 
                & \scriptsize Sanitizer    &                   & \scriptsize 99.20\% & \scriptsize \shortstack{\textcolor{darkgreen}{13.84\%} ($\downarrow$ 85.11\%)} & \scriptsize \shortstack{87.20\% ($\downarrow$ 2.82\%)} \\ \hline
\renewcommand{\arraystretch}{2.9} 
\multirow{2}{*}{\centering \scriptsize CIFAR10} & \scriptsize without-Sanitizer   & \multirow{2}{*}{\centering \scriptsize 92.12\%} & \scriptsize 98.88\% & \scriptsize \textcolor{red}{97.01\%} & \scriptsize \shortstack{88.06\% ($\downarrow$ 4.06\%)} \\ 
                & \scriptsize Sanitizer    &                   & \scriptsize 99.95\% & \scriptsize \shortstack{\textcolor{darkgreen}{12.22\%} ($\downarrow$ 84.79\%)} & \scriptsize \shortstack{87.18\% ($\downarrow$ 4.94\%)} \\ \hline
\renewcommand{\arraystretch}{2.9} 
\multirow{2}{*}{\centering \scriptsize CIFAR100} & \scriptsize without-Sanitizer   & \multirow{2}{*}{\centering \scriptsize 72.98\%} & \scriptsize 96.12\% & \scriptsize \textcolor{red}{95.42\%} & \scriptsize \shortstack{69.06\% ($\downarrow$ 3.92\%)} \\ 
                & \scriptsize Sanitizer    &                   & \scriptsize 94.40\% & \scriptsize \shortstack{\textcolor{darkgreen}{2.88\%} ($\downarrow$ 92.54\%)} & \scriptsize \shortstack{70.23\% ($\downarrow$ 2.75\%)} \\ \hline
\renewcommand{\arraystretch}{2.9} 
\multirow{2}{*}{\centering \scriptsize TinyImageNet} & \scriptsize without-Sanitizer   & \multirow{2}{*}{\centering \scriptsize 56.61\%} & \scriptsize 97.02\% & \scriptsize \textcolor{red}{97.89\%} & \scriptsize \shortstack{52.20\% ($\downarrow$ 4.41\%)} \\ 
                & \scriptsize Sanitizer    &                   & \scriptsize 100.00\% & \scriptsize \shortstack{\textcolor{darkgreen}{1.55\%} ($\downarrow$ 96.34\%)} & \scriptsize \shortstack{51.08\% ($\downarrow$ 5.53\%)} \\ \hline
\end{tabular}
}
\label{table_1}
\end{table*}

\begin{table*}[t]
\centering
\renewcommand{\arraystretch}{1.6} 
\scriptsize 
\caption{WDR, ASR, and ACC of Sanitizer compared to the without-Sanitizer under the conflicting scenarios across different datasets. Sanitizer maintains an exceptional WDR after sanitization and reduces the ASR to a level comparable to random guessing. The vanilla ACCs without any embedded backdoors or sanitization process are also reported.}
\resizebox{\textwidth}{!}{
\begin{tabular}{ccc>{\centering\arraybackslash}p{0.1\textwidth}>{\centering\arraybackslash}p{0.1\textwidth}>{\centering\arraybackslash}p{0.1\textwidth}}
\hline
\renewcommand{\arraystretch}{1.5} 
\multirow{2}{*}{\centering \scriptsize \textbf{Task}} & \multirow{2}{*}{\centering \scriptsize \textbf{Method}} & \multirow{2}{*}{\centering \scriptsize \textbf{Vanilla ACC}} & \multicolumn{3}{c}{\scriptsize \textbf{Scenario with Backdoor Conflicts}} \\ \cline{4-6} 
                &                   &                   & \scriptsize \textbf{WDR$\uparrow$} & \scriptsize \textbf{ASR$\downarrow$} & \scriptsize \textbf{ACC$\uparrow$} \\ \hline
\renewcommand{\arraystretch}{2.9} 
\multirow{2}{*}{\centering \scriptsize FMNIST} & \scriptsize without-Sanitizer   & \multirow{2}{*}{\centering \scriptsize 90.02\%} & \scriptsize 49.31\% & \scriptsize \textcolor{red}{97.83\%} & \scriptsize \shortstack{88.06\% ($\downarrow$ 1.96\%)} \\ 
                & \scriptsize Sanitizer    &                   & \scriptsize 100.00\% & \scriptsize \shortstack{\textcolor{darkgreen}{11.21\%} ($\downarrow$ 86.62\%)} & \scriptsize \shortstack{86.98\% ($\downarrow$ 3.04\%)} \\ \hline
\renewcommand{\arraystretch}{2.9} 
\multirow{2}{*}{\centering \scriptsize CIFAR10} & \scriptsize without-Sanitizer   & \multirow{2}{*}{\centering \scriptsize 92.12\%} & \scriptsize 62.30\% & \scriptsize \textcolor{red}{98.23\%} & \scriptsize \shortstack{87.50\% ($\downarrow$ 4.62\%)} \\ 
                & \scriptsize Sanitizer    &                   & \scriptsize 99.75\% & \scriptsize \shortstack{\textcolor{darkgreen}{10.55\%} ($\downarrow$ 87.68\%)} & \scriptsize \shortstack{86.92\% ($\downarrow$ 5.20\%)} \\ \hline
\renewcommand{\arraystretch}{2.9} 
\multirow{2}{*}{\centering \scriptsize CIFAR100} & \scriptsize without-Sanitizer   & \multirow{2}{*}{\centering \scriptsize 72.98\%} & \scriptsize 59.16\% & \scriptsize \textcolor{red}{94.72\%} & \scriptsize \shortstack{67.94\% ($\downarrow$ 5.04\%)} \\ 
                & \scriptsize Sanitizer    &                   & \scriptsize 96.12\% & \scriptsize \shortstack{\textcolor{darkgreen}{3.98\%} ($\downarrow$ 90.74\%)} & \scriptsize \shortstack{65.27\% ($\downarrow$ 7.71\%)} \\ \hline
\renewcommand{\arraystretch}{2.9} 
\multirow{2}{*}{\centering \scriptsize TinyImageNet} & \scriptsize without-Sanitizer   & \multirow{2}{*}{\centering \scriptsize 56.61\%} & \scriptsize 67.85\% & \scriptsize \textcolor{red}{95.44\%} & \scriptsize \shortstack{51.58\% ($\downarrow$ 5.03\%)} \\ 
                & \scriptsize Sanitizer    &                   & \scriptsize 98.71\% & \scriptsize \shortstack{\textcolor{darkgreen}{3.02\%} ($\downarrow$ 92.42\%)} & \scriptsize \shortstack{50.16\% ($\downarrow$ 6.45\%)} \\ \hline
\end{tabular}
}
\label{table_2}
\end{table*}

\noindent\textbf{Configuration and Hyperparameters}. Following the basic setup of existing works~\cite{wu2022toward,FL3,FL1,FL2,chow2023stdlens,shan2024geminio,li2023backdoorbox}, by default, we utilize FedAvg~\cite{mcmahan2017communication} as the aggregation backbone and simulate $K = 20$ participating clients, with half are malicious (embedding a malicious backdoor), while the other half are benign (embedding a benign watermark) for aggregation. Different values of $K$ are also provided for evaluating scalability. The number of communication rounds is set to 200. We employ stochastic gradient descent (SGD) optimization with $E = 10$ local epochs, a local initial learning rate of $lr = 0.01$, and a batch size of 128. The default poisoning rate for each client per round is fixed at 10\%. We specify that the default backdoor subnet size rate is set to 20\%. The rate of defense data is 0.05 of the whole dataset, and reverse engineering epochs is 2 for each round. We establish a unique pixel pattern of different shapes as the backdoor trigger pattern for each client and allocate a background image with a distinct solid color as the harmless environment. To ensure the rigor of the evaluation, all major experiments are conducted independently in triplicate, and we report the average value. We implement Sanitizer using PyTorch-2.4 and run all the experiments on a server with an AMD EPYC 7542 CPU (32 cores), 512 GB of memory, and an NVIDIA V100S GPU (32 GB), running Ubuntu 20.04 LTS on the CloudLab platform~\cite{zink2021open}.\\

\noindent\textbf{Outline}. We first evaluate the effectiveness of the proposed Sanitizer compared to the without-Sanitizer in Section~\ref{5-2} and Section~\ref{5-3}. Then we illustrate the performance of trigger recovery for certain clients across conflicting and non-conflicting scenarios (Section~\ref{5-4}). In Section~\ref{5-5}, we analyze the efficiency of Sanitizer against the baseline and further explore the scalability of Sanitizer, followed by ablation studies of Sanitizer’s key design components on efficiency. We show that Sanitizer is applicable across different neural architectures in Section~\ref{5-6}. Then, we investigate the impact of data heterogeneity on Sanitizer performance in Section~\ref{Non-iid}. In Section~\ref{Comparison_FedIPR}, we conduct an extensive analysis comparing Sanitizer with two additional methods. Finally, we show Sanitizer’s effectiveness and efficiency under different parameter settings in Section~\ref{5-7}.

\begin{table*}[t]
\centering
\normalsize
\caption{Two sets of visual original trigger pattern examples and their recovered versions by Sanitizer from CIFAR10 in the scenarios of non-conflicting and conflicting. In all cases, the corresponding trigger is successfully recovered, and each is assigned a unique input subspace as the harmless environment. This targeted assignment ensures that a high WDR is achieved only within its respective harmless environment.}
\begin{subtable}[t]{0.48\linewidth}
    \centering
    \caption{Scenario without trigger-output conflicts between clients.}
    \begin{tabular}{@{}>{\centering\arraybackslash}m{1.6cm}@{}>{\centering\arraybackslash}m{1.65cm}@{}>{\centering\arraybackslash}m{1.65cm}@{}>{\centering\arraybackslash}m{1.65cm}@{}>{\centering\arraybackslash}m{1.65cm}@{}}
    \toprule
    \normalsize Client \scriptsize (Target Class) & \normalsize Trigger & \normalsize Reversed Trigger & \small Harmless Environment & \normalsize WDR \\
    \midrule
    Client \textbf{1} (1) & \includegraphics[width=1.6cm, height=1.6cm]{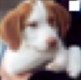} & \includegraphics[width=1.6cm, height=1.6cm]{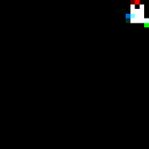} & \includegraphics[width=1.6cm, height=1.6cm]{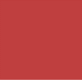} & 100.00\% \\
    \cmidrule(lr){1-5}
    Client \textbf{5} (2) & \includegraphics[width=1.6cm, height=1.6cm]{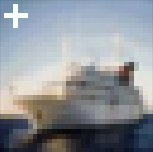} & \includegraphics[width=1.6cm, height=1.6cm]{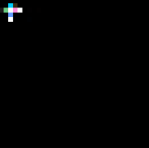} & \includegraphics[width=1.6cm, height=1.6cm]{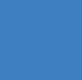} & 99.73\% \\
    \cmidrule(lr){1-5}
    Client \textbf{9} (3) & \includegraphics[width=1.6cm, height=1.6cm]{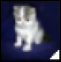} & \includegraphics[width=1.6cm, height=1.6cm]{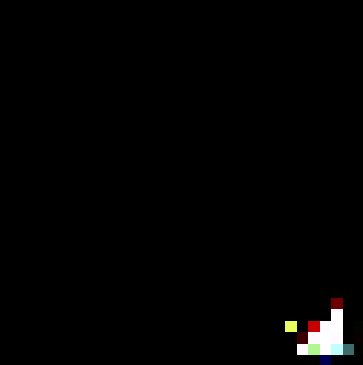} & \includegraphics[width=1.6cm, height=1.6cm]{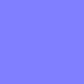} & 99.12\% \\
    \bottomrule
    \end{tabular}
    \label{table_non_conflict}
\end{subtable}%
\hfill
\begin{subtable}[t]{0.48\linewidth}
    \centering
    \caption{Scenario with trigger-output conflicts (e.g., Client 1 and 5).}
    \begin{tabular}{@{}>{\centering\arraybackslash}m{1.6cm}@{}>{\centering\arraybackslash}m{1.65cm}@{}>{\centering\arraybackslash}m{1.65cm}@{}>{\centering\arraybackslash}m{1.65cm}@{}>{\centering\arraybackslash}m{1.65cm}@{}}
    \toprule
    \normalsize Client \scriptsize (Target Class) & \normalsize Trigger & \normalsize Reversed Trigger &  \small Harmless Environment & \normalsize WDR \\
    \midrule
    Client \textbf{1} \textcolor{red}{(1)} & \includegraphics[width=1.6cm, height=1.6cm]{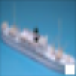} & \includegraphics[width=1.6cm, height=1.6cm]{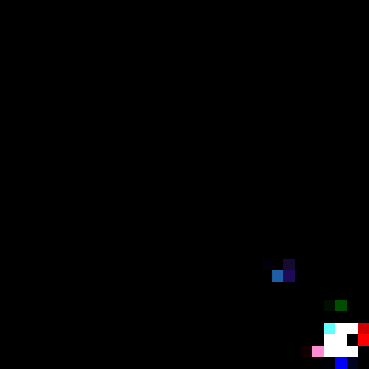} & \includegraphics[width=1.6cm, height=1.6cm]{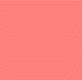} & 99.98\% \\
    \cmidrule(lr){1-5}
    Client \textbf{5} \textcolor{red}{(2)} & \includegraphics[width=1.6cm, height=1.6cm]{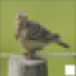} & \includegraphics[width=1.6cm, height=1.6cm]{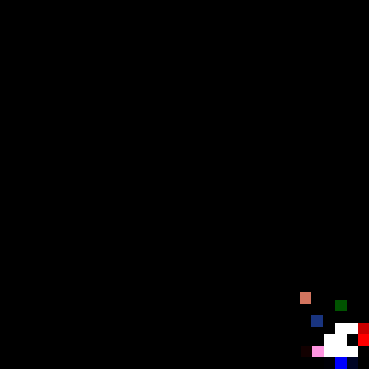} & \includegraphics[width=1.6cm, height=1.6cm]{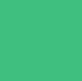} & 99.93\% \\
    \cmidrule(lr){1-5}
    Client \textbf{9} (3) & \includegraphics[width=1.6cm, height=1.6cm]{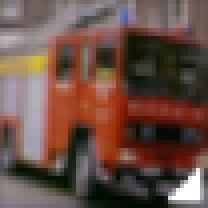} & \includegraphics[width=1.6cm, height=1.6cm]{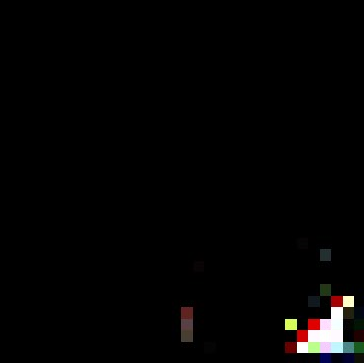} & \includegraphics[width=1.6cm, height=1.6cm]{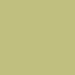} & 98.69\% \\
    \bottomrule
    \end{tabular}
    \label{table_conflict}
\end{subtable}
\label{table_combined}
\end{table*}

\subsection{Effectiveness in Non-conflicting Scenario}
\label{5-2}
We first evaluate Sanitizer compared to without-Sanitizer (i.e., existing backdoor-based client-side watermarked FL without Sanitizer) in terms of watermark effectiveness (WDR), harmfulness (ASR), and fidelity (ACC) under the non-conflicting scenario we set. Table~\ref{table_1} reports the average results obtained from evaluating the final FL-trained model. Besides, a vanilla ACC of an FL-trained model without any embedded backdoors or sanitization process is also provided as a reference for comparisons.

On one hand, it can be observed that the WDR, averaged across all benign clients, of the model without Sanitizer is high, all exceeding 96\%, while the ASR, averaged across all malicious clients, also exhibits a substantial level of at least 95\% for all datasets. Both watermarks and harmful backdoors remain active in the unprotected watermarked FL-trained model. These results indicate that malicious clients can effortlessly implant malicious backdoors into the model for harmful purposes during the backdoor-based client-side watermarking process, underscoring the inherent vulnerability and susceptibility of the FL-trained model in this context. 

In contrast, after applying Sanitizer to FL, the WDR consistently remains at least 94\% across four datasets, which is high enough for ownership and contribution verification. Furthermore, the ASR is dramatically reduced to a random guess level, with a sharp drop compared to without-Sanitizer. In particular, the ASR for all four tasks demonstrates a reduction of over 84\%. Sanitizer causes a slight decrease in the model's ACC compared to the vanilla ACC, similar to the behavior observed in without-Sanitizer. Taken together, these results reveal several outcomes achieved by Sanitizer. First, Sanitizer successfully transforms the watermarks into a harmless state, thus confirming the effectiveness of our novel harmless watermarks in FL. Simultaneously, Sanitizer effectively mitigates the threats of malicious backdoors, demonstrating its ability to precisely identify and prune the backdoor subnets from the client-submitted models. Specifically, this prevents backdoors from being triggered on natural queries in harmful ways, thus safeguarding the model's integrity. Furthermore, Sanitizer has a negligible impact on the fidelity of the FL-trained model. Although there is a slight drop in ACC hovering around 5\%, the trade-off is acceptable in our context in light of the substantial reduction in ASR. 

Collectively, Sanitizer offers strong effectiveness in making watermarks harmless in FL and is widely applicable and not restricted to specific datasets.

\subsection{Effectiveness in Conflicting Scenario}
\label{5-3}
In the scenario with backdoor conflicts, we introduce multiple pairs of watermark conflicts among the benign clients. Specifically, we simulate watermark conflicts by employing identical triggers with different target classes, thereby creating deliberate conflicts in watermark verification. For example, both Clients 1 and 5 use a white square as the trigger; however, Client 1 assigns Class 1 as the target label, while Client 5 assigns Class 2, which will be visualized in Section~\ref{5-4}. As presented in Table~\ref{table_2}, without-Sanitizer is unable to achieve a satisfactory WDR, while the corresponding ASR remains dangerously high. This confirms that conflicts compromise the effectiveness of the watermark by making them unlearnable, leading to ambiguous contribution claims. Meanwhile, the backdoor attack remains highly effective. After Sanitizer conducts the harmless transformation for every backdoor independently, the WDR improves to an impressive rate exceeding 96\%, which is sufficient for contribution verification. More importantly, the ASR experiences a marked reduction of at least 86\%, and ACC also remains competitive at a considerable level, with a moderate ACC drop of around 5\% induced by Sanitizer. The experimental results across four datasets are consistent with those observed in the non-conflicting setting. Sanitizer reduces the ASR to a near-random guess level while maintaining a high WDR and a robust ACC in the conflicting scenario. What stands out from this experiment is that Sanitizer also effectively transforms the conflicting backdoors into harmless and conflict-free ones while mitigating the harmful backdoor effect and preserving the main functionality in the scenario with watermark conflicts.

To distill the findings, Sanitizer consistently demonstrates its general effectiveness in both conflict-free and conflict-present scenarios. Why is it effective? The key lies in its ability to successfully and systematically reverse the embedded trigger pattern in each client-submitted model during the FL process and eventually confine them to their own harmless, client-specific input subspace. We will provide a more in-depth explanation via visualization in the following subsection.

\subsection{Visual Examples of Trigger Inversion}
\label{5-4}
The previous results demonstrate the overall effectiveness of Sanitizer. In this subsection, we focus on specific clients and employ trigger visualization to further validate the effectiveness of Sanitizer in trigger pattern recovery. All hyperparameters and configurations are maintained in default settings. Table~\ref{table_combined} provides two visual example sets of the original trigger patterns (2nd column) and their eventually recovered versions (3rd column) by Sanitizer from CIFAR10. In the scenario without trigger-output conflicts (Table~\ref{table_non_conflict}), we apply three distinct, commonly used patch-like patterns to the clean data for three participating clients (Client 1, Client 5, and Client 9), setting their desired outputs to different classes (1, 2, and 3), respectively. As we can see, in all cases, the corresponding backdoor trigger is successfully recovered, regardless of whether its real-world purpose is benign or malicious. Specifically, Client 1 achieves a notable trigger recovery and reaches a WDR of 100.00\% after Sanitizer relearns the reversed trigger-output mapping rule into the unique and harmless input subspace (4th column). Client 5 and Client 9 demonstrate comparable recovery performance with WDR of 99.73\% and 99.12\%, respectively. 

\begin{table*}[t]
\centering
\renewcommand{\arraystretch}{1.2}
\small
\caption{Efficiency comparison of the baseline and Sanitizer on CIFAR10 using ResNet18 and MobileNetV3. Sanitizer demonstrates significantly improved efficiency compared to the baseline on two architectures, with a substantial reduction in both resource and time consumption while maintaining comparable effectiveness to the baseline.}
\begin{subtable}[t]{0.48\textwidth}
    \centering
    \begin{tabular}{@{}>{\centering\arraybackslash}p{2.2cm}@{}>{\centering\arraybackslash}p{1.9cm}@{}>{\centering\arraybackslash}p{1.5cm}@{}>{\centering\arraybackslash}p{2.4cm}@{}}
    \toprule
    \multirow{2}{*}{\textbf{Method}} & \multicolumn{3}{c}{\textbf{CIFAR10 \& ResNet18}} \\
    \cmidrule(lr){2-4}
     & \textbf{GPU\_Mem} & \textbf{Time} & \textbf{Precision} \\ 
    \midrule
    \multirow{3}{*}{\normalsize \shortstack{Baseline}} & \multirow{3}{*}{\normalsize \textcolor{red}{1322MB}} & \multirow{3}{*}{\normalsize \textcolor{red}{142.50s}} & WDR:99.87\% \\ 
     &  &  & ASR:12.98\% \\
     &  &  & ACC:88.70\% \\
    \cmidrule(lr){1-4}
    \multirow{3}{*}{\normalsize \shortstack{Sanitizer}} & \multirow{3}{*}{\shortstack{\normalsize \textcolor{darkgreen}{194MB} \\ ($\downarrow$ 85.32\%)}} & \multirow{3}{*}{\shortstack{\normalsize \textcolor{darkgreen}{28.12s} \\ ($\uparrow$ 5x)}} & WDR:99.95\% \\ 
     &  &  & ASR:12.22\% \\
     &  &  & ACC:87.18\% \\
    \bottomrule
    \end{tabular}
    \label{table_cifar10_comparison_1}
\end{subtable}%
\hfill 
\begin{subtable}[t]{0.48\textwidth}
    \centering
    \begin{tabular}{@{}>{\centering\arraybackslash}p{2.2cm}@{}>{\centering\arraybackslash}p{1.9cm}@{}>{\centering\arraybackslash}p{1.5cm}@{}>{\centering\arraybackslash}p{2.4cm}@{}}
    \toprule
    \multirow{2}{*}{\textbf{Method}} & \multicolumn{3}{c}{\textbf{CIFAR10 \& MobileNetV3}} \\
    \cmidrule(lr){2-4}
     & \textbf{GPU\_Mem} & \textbf{Time} & \textbf{Precision} \\ 
    \midrule
    \multirow{3}{*}{\normalsize \shortstack{Baseline}} & \multirow{3}{*}{\normalsize \textcolor{red}{273MB}} & \multirow{3}{*}{\normalsize \textcolor{red}{84.88s}} & WDR:99.06\% \\ 
     &  &  & ASR:13.22\% \\
     &  &  & ACC:87.34\% \\
    \cmidrule(lr){1-4}
    \multirow{3}{*}{\normalsize \shortstack{Sanitizer}} & \multirow{3}{*}{\shortstack{\normalsize \textcolor{darkgreen}{65MB}\\ ($\downarrow$ 76.19\%)}} & \multirow{3}{*}{\shortstack{\normalsize \textcolor{darkgreen}{17.46s} \\ ($\uparrow$ 4.8x)}} & WDR:100.00\% \\ 
     &  &  & ASR:9.75\% \\
     &  &  & ACC:84.66\% \\
    \bottomrule
    \end{tabular}
    \label{table_tiny_comparison_1}
\end{subtable}
\label{table_combined_efficiency}
\end{table*}

On the other hand, Table~\ref{table_conflict} presents results in the scenario where trigger-output conflicts exist, particularly between Client 1 and Client 5, both of whom use an identical trigger pattern (a 4$\times$4 white square in the bottom-right corner) while aiming for different target classes (label 1 and label 2). We observe that Sanitizer is insensitive to the presence of backdoor conflicts, as the trigger patterns recovered by Sanitizer continue to be feasible, and the WDR consistently remains high, with Client 1, Client 5, and Client 9 exhibiting 99.98\%, 99.93\%, and 98.69\%, respectively. This effectiveness is attributed to two facts. First, each client is processed independently before aggregation during the FL process. Second, the final harmless input subspace is client-specific, effectively resolving the conflict issue. Furthermore, we observe that recovering the trigger from an FL-trained model without Sanitizer appears to be not feasible. Additional results can be found in the Appendix~\ref{appendix_A5}. Besides, we also report the success rate when only the harmless background is provided (i.e., no trigger is added) during the verification stage. The result is 0\%, outputting a randomly fixed label different from the target class. This shows that the trigger’s behavior relies on both the subspace (harmless background) and the trigger content, as the model learns their binding relation during the harmless relearning stage. Using either alone cannot achieve the intended effect in our context.

\textbf{Analysis on Effectiveness of Trigger Inversion}. First, why can the small extracted backdoor subnet stemming from the client-submitted model be used for trigger inversion? Thanks to the clean unlearning part (Section~\ref{sec: Identify Backdoor Subnets}), we identify a small set of backdoor-related units within the original client-submitted model, which constitute a backdoor subnet. This backdoor subnet fully encapsulates backdoor knowledge while preserving most of the backdoor functionality, as demonstrated by the low ACC and high WDR of the subnet. For instance, on the CIFAR10 using ResNet18, the backdoor subnet achieves an average ACC of 14.60\% and WDR of 99.98\%. When applying the trigger recovery and backdoor detection on the extracted backdoor subnet, one can more easily expose the potential backdoor target and enhance the quality of the recovered trigger pattern. Second, why is the round-spread trigger recovery feasible? In each round, the models at the client-side are trained on a sanitized global model from the previous round, and they undergo continuous backdoor attacks for either benign or harmful purposes in this round. The backdoor task exhibits rapid convergence, achieving a high WDR or ASR greater than 99\% quickly during local training. Upon receiving each client-submitted model, Sanitizer utilizes the results of reverse engineering from the previous round for that client as the initial values for the reverse optimization in the current round, ultimately facilitating the successful reconstruction of the backdoor trigger for each client. Moreover, a key aspect of Sanitizer's effectiveness is that the trigger from each client is confined to its own harmless and client-specific input subspace (i.e., being effective only in the harmless environment), which is directly attributed to the success of the trigger inversion step.

\subsection{Efficiency and Scalability Studies}
\label{5-5}

\subsubsection{Efficiency of Sanitizer}
\label{Efficiency of Sanitizer}
As both Sanitizer and the baseline (described in Section~\ref{sec:intro}) aim to achieve the identically expected effectiveness, we primarily evaluate the differences in their efficiency during the FL process, which is a critical concern in real-world FL applications. To this end, we evaluate the efficiency of the two methods on two different architectures, with the other settings kept identical to those in Section~\ref{5-3}. We use the following primary metrics: (i) the GPU memory utilized GPU\_Mem (MB) for processing one client-submitted model during server-side operations, and (ii) the average runtime Time (s) of server-side operations per round during FL.

\textbf{Results on ResNet18}. Table~\ref{table_combined_efficiency} reports the results. What most strikingly emerges from it is that despite both approaches achieving considerable effectiveness, Sanitizer exhibits significantly less overhead in terms of resource consumption and runtime compared to the baseline. Specifically, the baseline consumes 1,322 MB of GPU memory for each client-submitted model and takes 142.50 seconds for server-side execution per round. On the contrary, Sanitizer drastically reduces the GPU memory usage to 194 MB with an 85\% reduction, which suggests that in each round, the server can process more client-submitted models in parallel. Moreover, the runtime optimization achieved by Sanitizer per round over the baseline is also particularly significant, with a 5x speedup in execution time from 142.50 to 28.12 seconds, highlighting the efficiency gains provided by Sanitizer. With the significantly reduced resource and time consumption, Sanitizer still maintains non-trivial effectiveness, closely aligning with the performance of the baseline.

\textbf{Results on MobileNetV3}. The architectures in the MobileNet family are inherently designed with fewer parameters and lower computational cost compared to those in the ResNet family. As shown in Table~\ref{table_combined_efficiency}, the experimental results consistently display similar behaviors. Remarkably, Sanitizer shows a drop of over 75\% in GPU memory consumption and an approximately 5x speedup in runtime per round compared to the baseline. Meanwhile, as illustrated, Sanitizer also achieves the intended sanitization results. Together these results indicate Sanitizer's suitability for lightweight neural architectures, which are commonly used in resource-constrained environments where computational and memory limitations are critical considerations.

In this subsection, we derive the important insight that Sanitizer demonstrates a significant efficiency advantage while maintaining comparable effectiveness, harmlessness, and fidelity. This is attributed to the fact that Sanitizer performs the round-spread trigger recovery on the extracted small backdoor subnet, two improvements introduced to accelerate the entire FL period. Moreover, the reduction in the GPU memory consumption makes Sanitizer particularly attractive for IP protection in resource-constrained FL environments. Interestingly, this may reflect the scalability of Sanitizer, as already shown in Figure~\ref{intro_fig1} of Section~\ref{sec:intro}, prompting us to further investigate this characteristic, which will be discussed in detail in subsection~\ref{5-2-3-1}.

\begin{figure}[t]
\centering
\includegraphics[width=0.95\columnwidth]{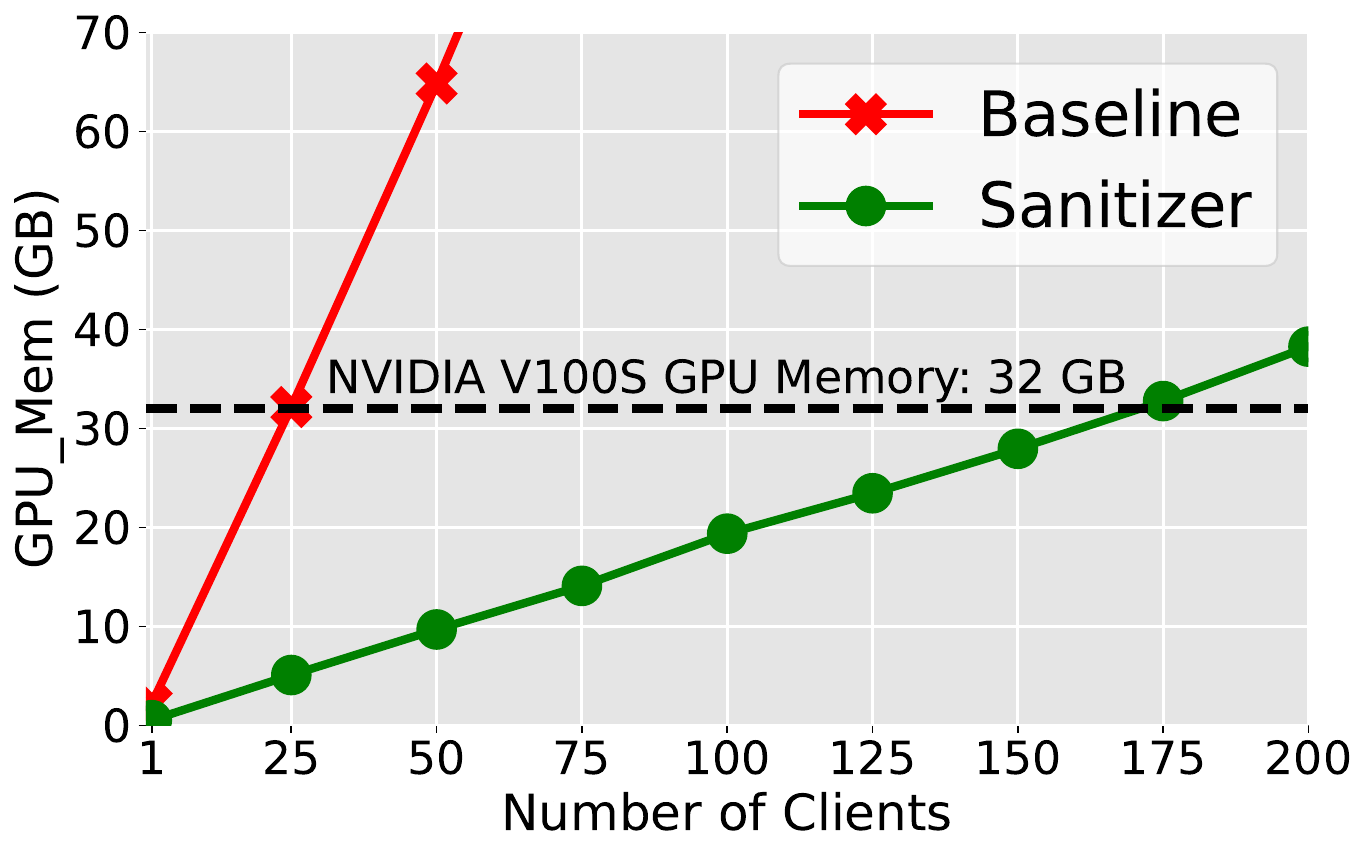} 
\caption{Sanitizer offers significantly improved scalability. It keeps GPU memory consumption (green) steadily growing but consistently below the server's total GPU capacity of 32 GB (black) as the number of clients increases during a 10-round FL on CIFAR10 with ResNet18.}
\label{memory_cost_fig1}
\end{figure}

\subsubsection{Scalability of Sanitizer}
\label{5-2-3-1}
To evaluate the scalability of Sanitizer, we examine the total server-side time consumption of two methods during the whole FL process as the number of participating clients increases over a 10-round FL on CIFAR10 using ResNet18. Figure~\ref{intro_fig1} clearly showcases the scalability advantage of Sanitizer when compared to the baseline. Specifically, for the baseline, the time cost increases rapidly, showing near-exponential growth beyond 50 clients. Notably, with 75 clients, the baseline requires over 4,500 seconds to complete, which exceeds the acceptable range we set, and this steep upward trend persists as the number of clients continues to rise. Such scaling behavior suggests the baseline struggles to maintain practical efficiency in large-scale FL scenarios, as the time required to complete sanitization skyrockets when more clients are added. In contrast, Sanitizer exhibits robust scalability across the entire range of clients tested. As shown by the green line, Sanitizer’s time complexity remains consistently low (below 1,000 seconds) when scaling from 1 to 150 clients and shows only a moderate linear increase as the number of clients approaches 200. Under the fixed server's GPU memory, the divergence between the two methods, which initially is fivefold, continues to increase progressively. The stark contrast in time consumption suggests that the improved designs we introduce into Sanitizer allow it to effectively alleviate the severe server-side computation overhead observed in the baseline, making it far more well-suited for real-world deployment in large-scale FL settings, where hundreds or even thousands of clients may be commonplace. Additionally, we believe that the scalability of Sanitizer is also represented in its optimization of GPU resource consumption.

\begin{table}[t]
  \centering
  \normalsize
  \renewcommand{\arraystretch}{1.4}  
  \setlength{\tabcolsep}{6.6pt}  
  \caption{Ablation studies showing the impact of Sanitizer's two key individual components on efficiency for understanding the contribution of different proposed designs.}
  \begin{tabular}{lcc}
    \hline
    & \textbf{GPU\_Mem} & \textbf{Time} \\
    \hline
    Baseline & 1322MB & 142.50s \\
    (a) only Subnet Extraction & 196MB & 113.44s \\
    (b) only Round Spreading & 1318MB & 69.61s \\
    \hline
    \rule{0pt}{3ex}(a+b) Sanitizer & 194MB & 28.12s \\
    \hline
  \end{tabular}
  \label{tab_ablation_study}
\end{table}

Figure~\ref{memory_cost_fig1} illustrates the comparison of GPU memory consumption between the two methods under the same experimental settings. The baseline exhibits a sharp increase in GPU memory consumption as the number of clients increases. By the time it exceeds 25 clients, the baseline has already surpassed the threshold of 32 GB, a limit for the maximum capacity of the server's GPU memory in our setting, indicating that a maximum of 25 client-submitted models can be processed concurrently on the server-side. In comparison, Sanitizer is capable of scaling the number of concurrently processed clients beyond 175 while remaining below the threshold, without imposing a significant burden on the server. This is essentially due to the reduced GPU memory consumed by a single backdoor subnet, making Sanitizer well-suited for large-scale FL applications with constrained server-side resources. To sum up, these results provide a compelling argument for the deployment of Sanitizer in real-world FL environments with a large number of clients.

\begin{figure*}[t]
\centering
\includegraphics[width=0.999\textwidth]{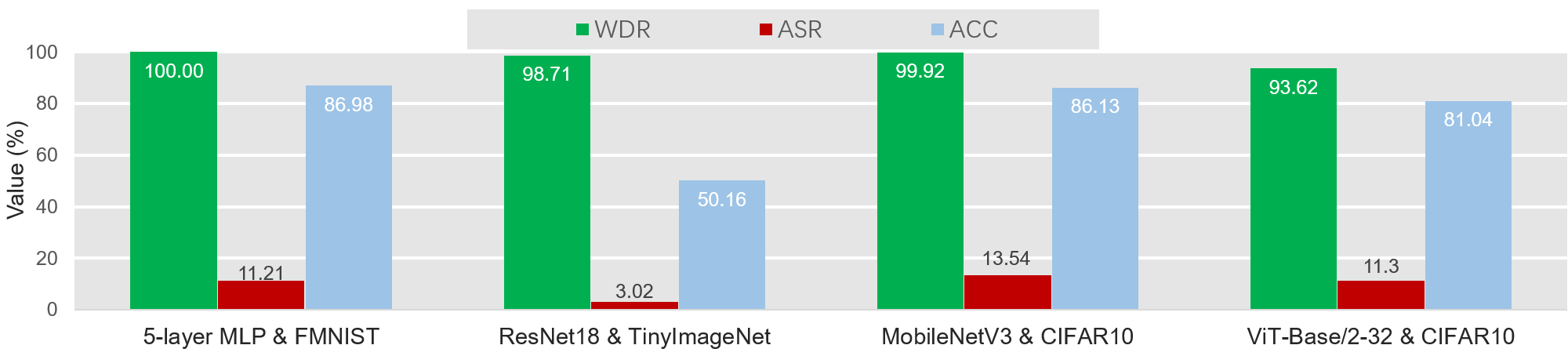}
\caption{WDR, ASR, and ACC of Sanitizer across different architectures under the practical scenario with conflicts.}
\label{sub_fig_precision_diff_archs}
\end{figure*}

\subsubsection{Ablation Studies}
\label{ablation_study}
To verify the individual functionality of each key component in Sanitizer and the importance of utilizing them for improving the efficiency, we conduct ablation studies by removing the Subnet Extraction component and Round-spread Optimization component in Sanitizer at a time in default experimental settings. As illustrated in Table~\ref{tab_ablation_study}, compared to the baseline, the variant with Subnet Extraction component only (a) cuts the occupied memory per model by approximately 85\%, while achieving about a 20\% improvement in per-round runtime over the baseline. If we apply Round-spread optimization without extracting subnets (b), it achieves a drastic reduction in runtime by over 51\% while maintaining similar memory consumption at 1,318 MB. The final row (a+b) demonstrates the combined effect of both optimizations enabled in our complete Sanitizer, achieving superior performance in both memory usage and runtime compared to either component alone or the baseline. We can derive the following conclusions: (i) Subnet Extraction component is the decisive contributor to GPU memory efficiency by confining the backdoor analysis to a small subnet. Round-spread component alone does not alleviate memory load; (ii) in terms of reducing time overhead, both components demonstrate effectiveness, with the Round-spread component performing better by distributing the inversion optimization across multiple rounds, chiefly contributing to runtime efficiency; (iii) the results demonstrate that employing both components yields the best of both worlds while providing complementary benefits, further indicating their necessity for the success of Sanitizer.


\begin{table}[t]
  \centering
  \normalsize
  \renewcommand{\arraystretch}{1.4}  
  \setlength{\tabcolsep}{9.0pt}  
  \caption{Effectiveness comparison of Sanitizer following the Dirichlet Distribution across different concentration parameters $\alpha$, showing Sanitizer's robust performance across different degrees of data heterogeneity.}
  \begin{tabular}{cccc}
    \hline
    Concentration & \textbf{WDR $\uparrow$} & \textbf{ASR $\downarrow$} & \textbf{ACC $\uparrow$}\\
    \hline
    $\alpha = 0.7$ & 98.61\% & 17.90\%  &  77.21\% \\
    $\alpha = 0.8$ & 99.22\% & 14.23\%  &  83.28\% \\
    $\alpha = 0.9$ & 97.13\% & 12.50\%  &  84.63\% \\
    \hline
    \rule{0pt}{3ex}(IID) Sanitizer & 99.95\% & 12.22\% & 87.18\% \\
    \hline
  \end{tabular}
  \label{tab_non-iid}
\end{table}

\subsection{Generalizability of Sanitizer on Different Neural Architectures}
\label{5-6}

We also observe that Sanitizer remains effective across various neural architectures. Figure~\ref{sub_fig_precision_diff_archs} reports the effectiveness of Sanitizer in the scenario with backdoor conflicts, which represents more realistic conditions. Whether applied to traditional fully connected architectures like MLP, classical convolutional architectures such as ResNet18, or lightweight architectures like MobileNetV3, Sanitizer consistently achieves a near-perfect WDR of over 98\%, demonstrating its ability to reliably confine the watermarks to a harmless and client-specific input subspace and enable the triggering of counter-intuitive model behavior for contribution demonstration in harmless, non-conflicting ways. The ASR is reduced from a relatively elevated level to that of random guessing, underscoring that Sanitizer can effectively neutralize malicious backdoors. Additionally, the model's fidelity is largely preserved, with the ACC typically remaining high. 

We enrich our evaluation with ViT~\cite{dosovitskiy2020image}, even though Transformer-based models are less common in FL due to their computational burden on client devices. Specifically, the core component of ViT, the multi-head self-attention layer, divides the parameter space into multiple heads, where each head specializes in capturing a specific aspect of knowledge. Leveraging this, Sanitizer can identify and extract backdoor-related heads, the associated head channels, and the corresponding hidden dimensions in the MLP module. As shown in Figure~\ref{sub_fig_precision_diff_archs}, Sanitizer works similarly well with a 93.62\% success rate for harmless watermark verification and an 11.30\% success rate for malicious attacks. Taken together, this analysis further reveals that Sanitizer is a versatile and architecture-agnostic method that can be effectively applied to different neural architectures in FL.

\begin{figure*}[t]
    \centering
    \begin{subfigure}[t]{0.24\textwidth}
        \centering
        \includegraphics[width=\textwidth]{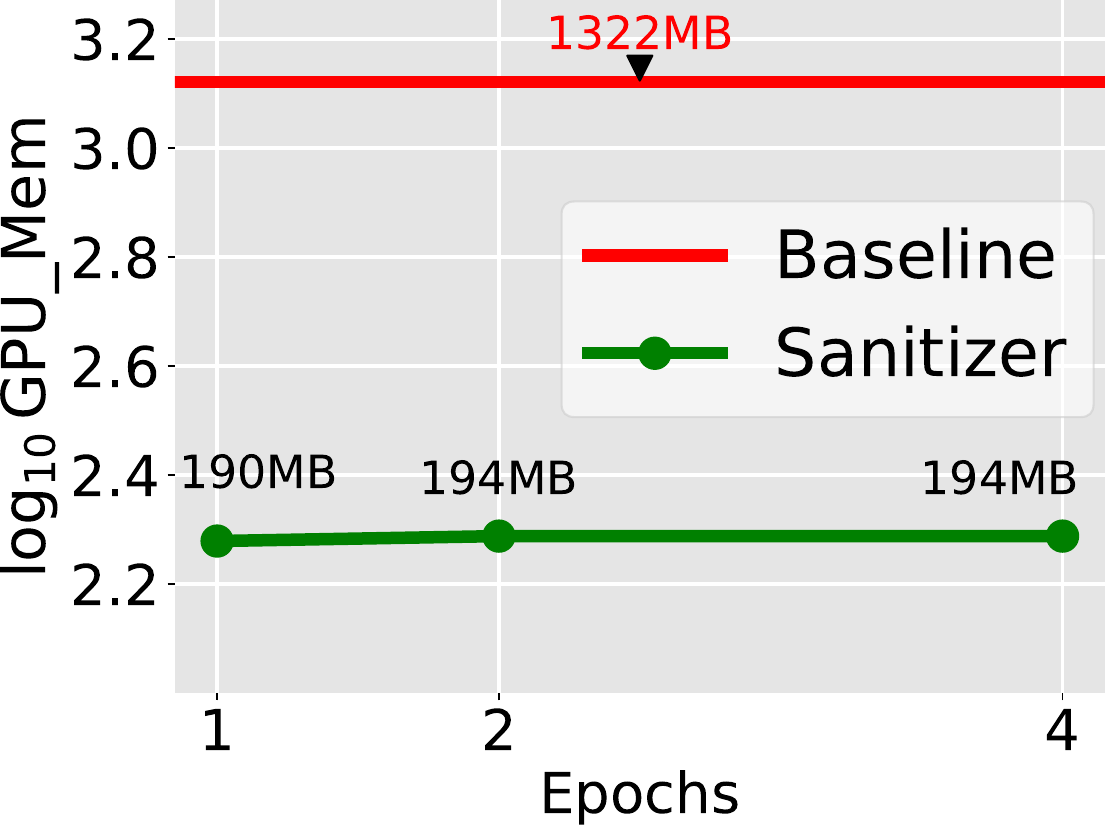}
        \label{sub_fig_variant_epochs:a}
    \end{subfigure}
    \hfill
    \begin{subfigure}[t]{0.24\textwidth}
        \centering
        \includegraphics[width=\textwidth]{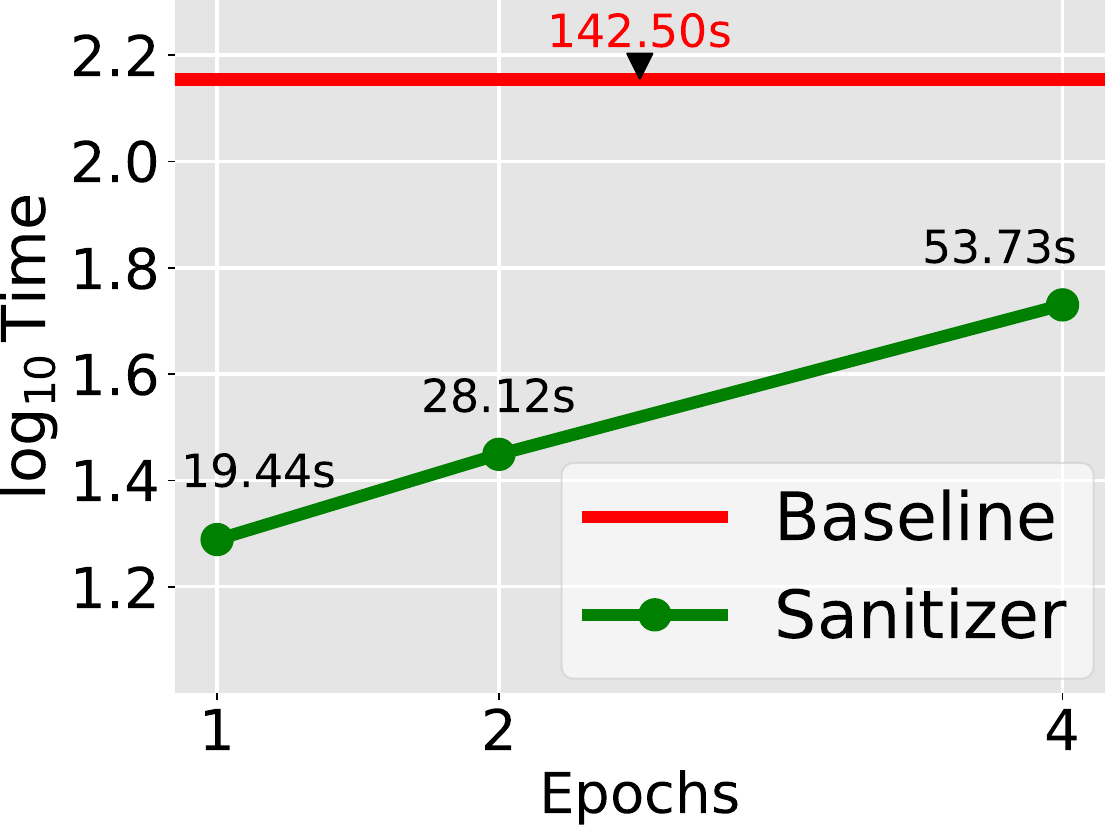}
        \label{sub_fig_variant_epochs:b}
    \end{subfigure}
    \hfill
    \begin{subfigure}[t]{0.24\textwidth}
        \centering
        \includegraphics[width=\textwidth]{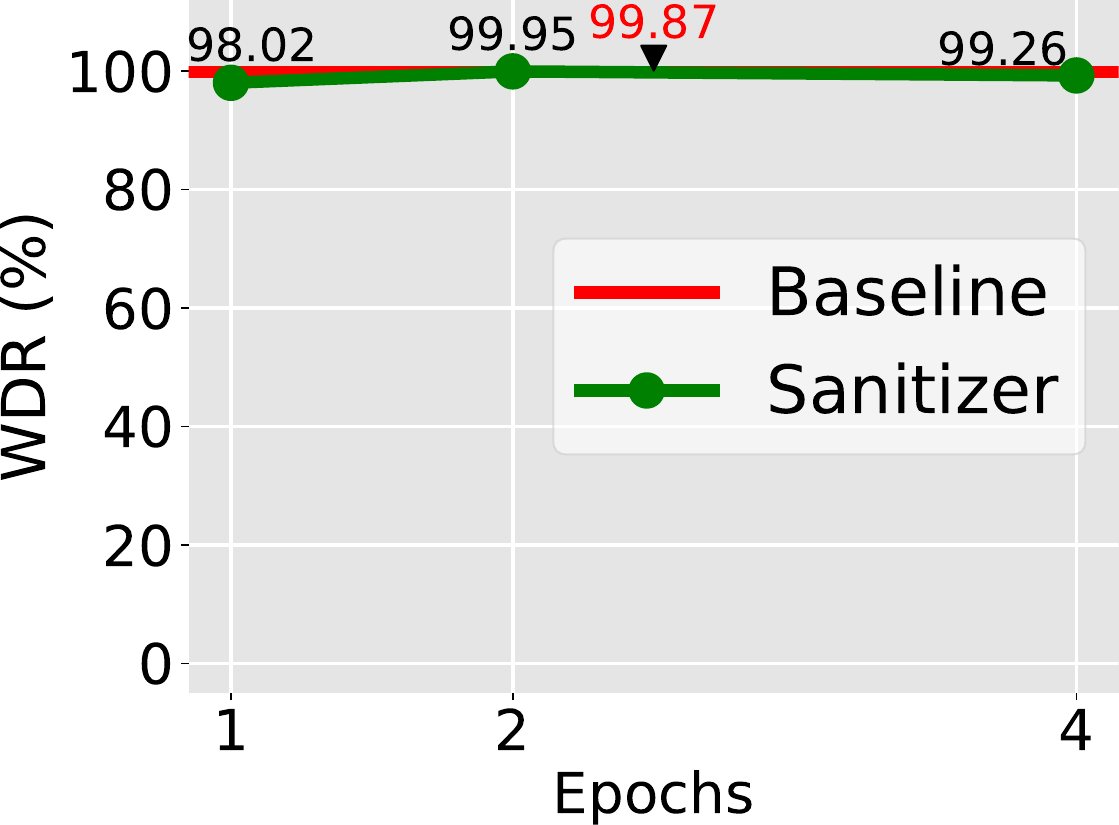}
        \label{sub_fig_variant_epochs:c}
    \end{subfigure}
    \hfill
    \begin{subfigure}[t]{0.24\textwidth}
        \centering
        \includegraphics[width=\textwidth]{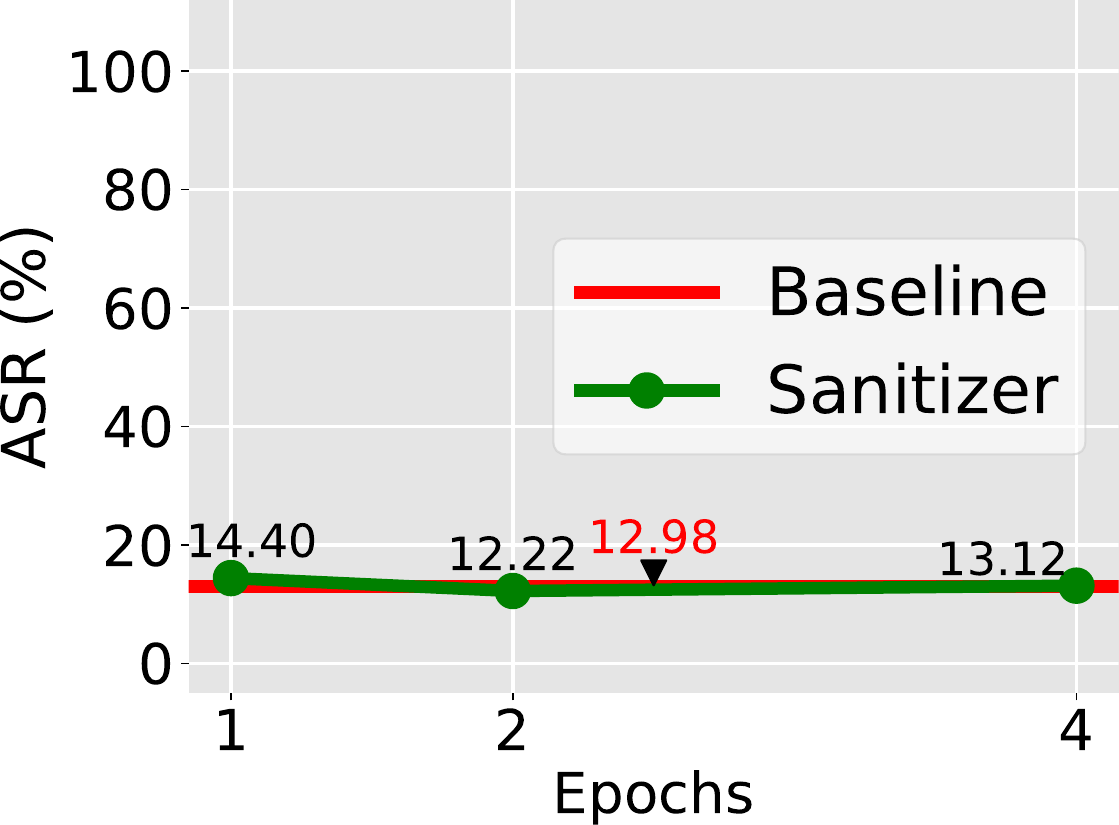}
        \label{sub_fig_variant_epochs:d}
    \end{subfigure}
    \caption{Efficiency and effectiveness metrics across different settings of the round-spread reverse engineering epochs.}
    \label{variant_epochs_all}
\end{figure*}

\begin{figure*}[t]
    \centering
    \begin{subfigure}[t]{0.24\textwidth}
        \centering
        \includegraphics[width=\textwidth]{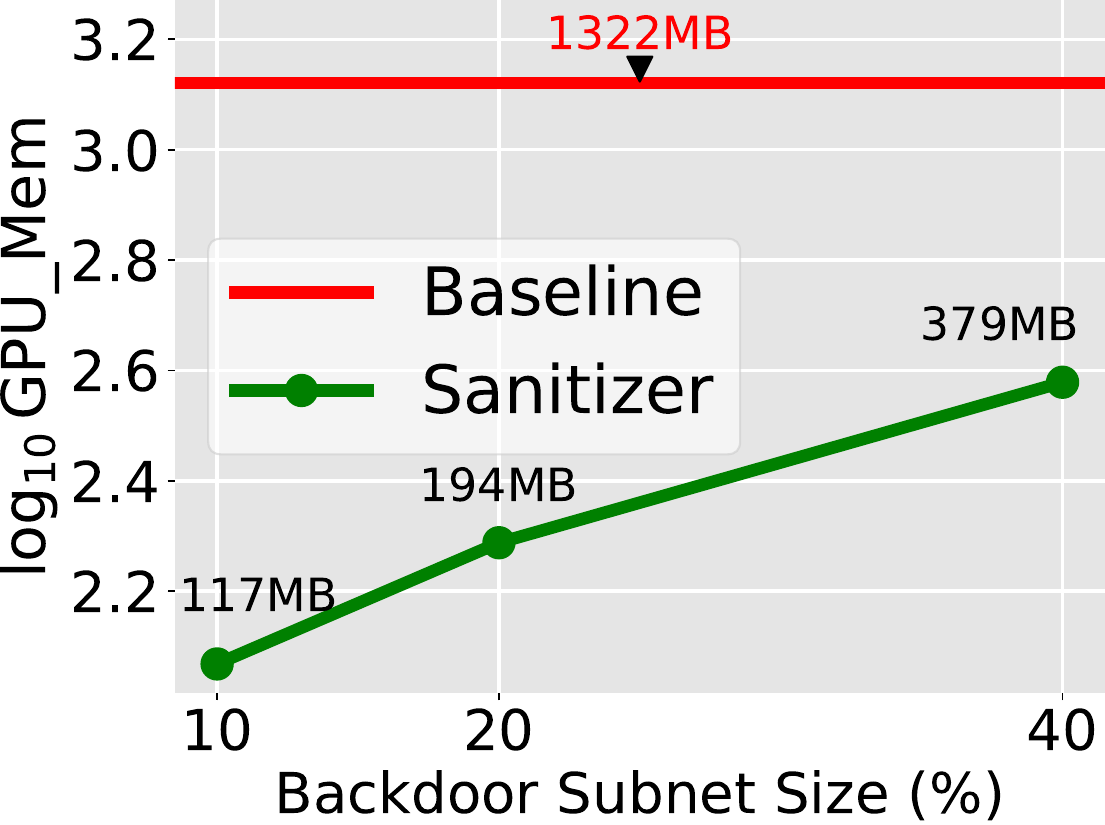}
        \label{sub_fig_variant_subnet:a}
    \end{subfigure}
    \hfill
    \begin{subfigure}[t]{0.24\textwidth}
        \centering
        \includegraphics[width=\textwidth]{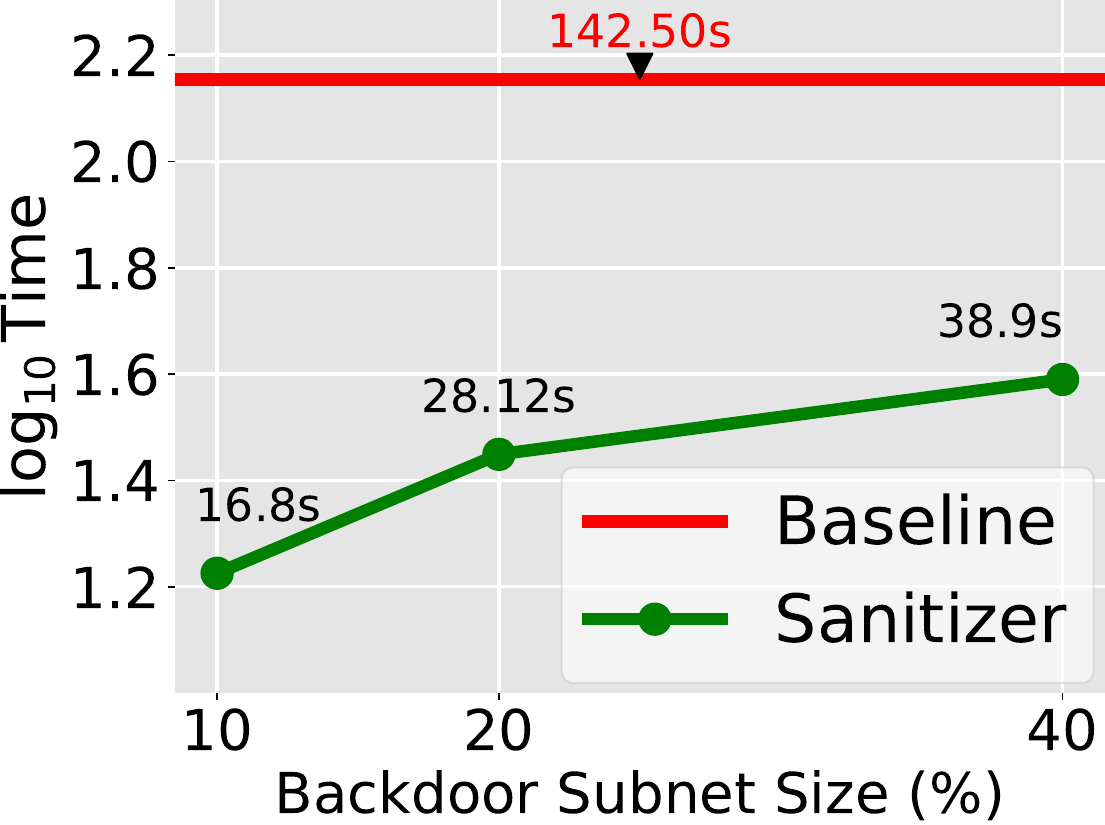}
        \label{sub_fig_variant_subnet:b}
    \end{subfigure}
    \hfill
    \begin{subfigure}[t]{0.24\textwidth}
        \centering
        \includegraphics[width=\textwidth]{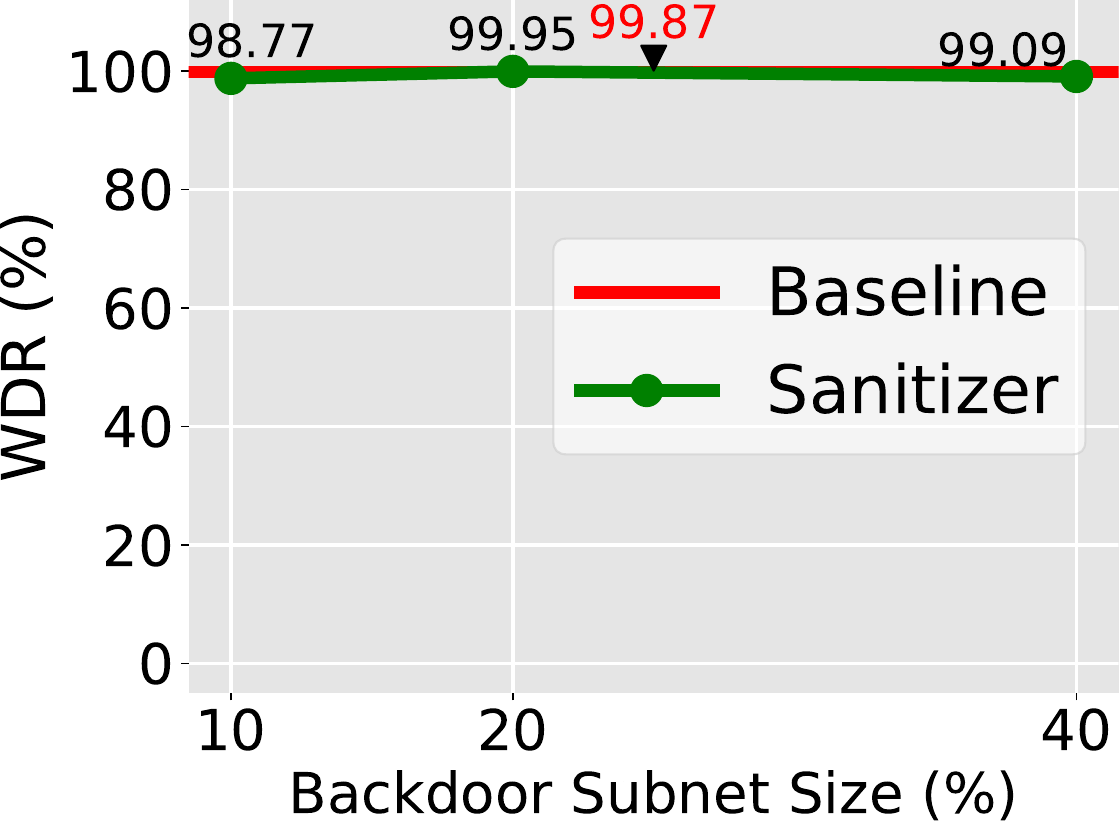}
        \label{sub_fig_variant_subnet:c}
    \end{subfigure}
    \hfill
    \begin{subfigure}[t]{0.24\textwidth}
        \centering
        \includegraphics[width=\textwidth]{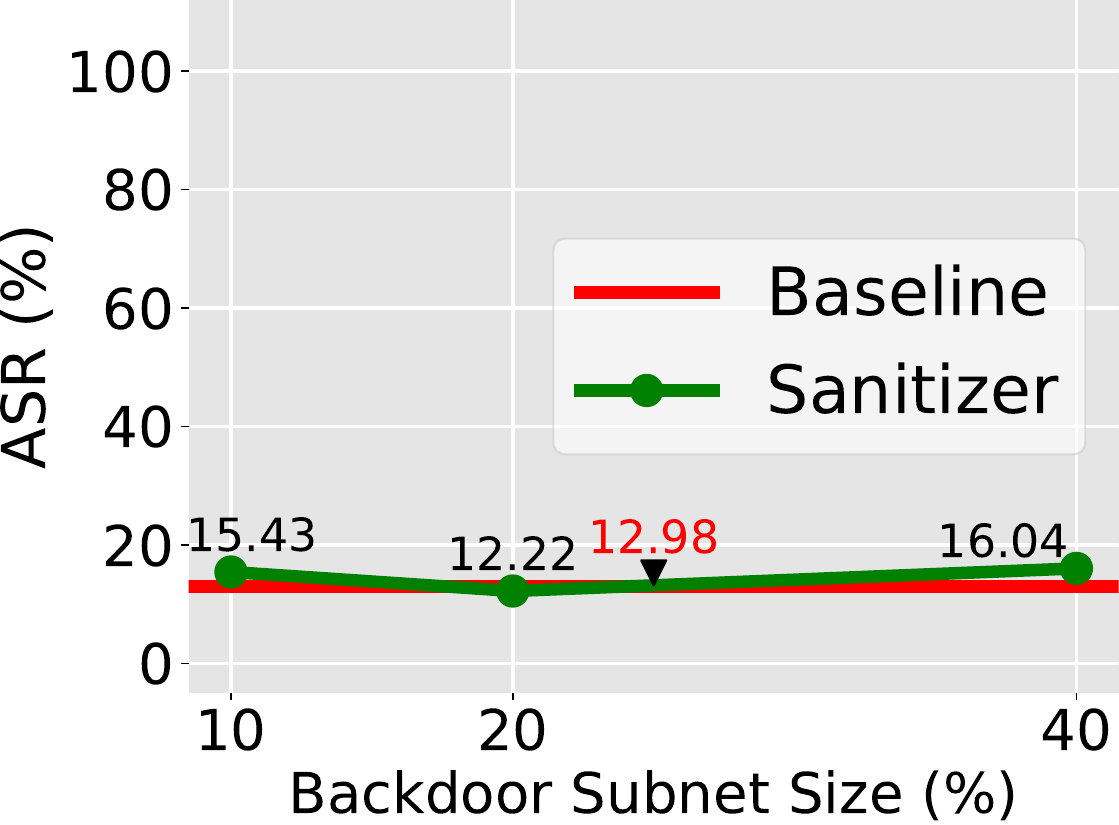}
        \label{sub_fig_variant_subnet:d}
    \end{subfigure}
    \caption{Efficiency and effectiveness metrics across different settings of the backdoor subnet size (\%).}
    \label{variant_subnet_all}
\end{figure*}

\subsection{Evaluation of Sanitizer under Non-IID Setting}
\label{Non-iid}
In a more realistic setting for FL, client data distributions frequently exhibit non-independent and identically distributed (non-IID) characteristics, reflecting the heterogeneity of local data among participating clients. To study whether data heterogeneity influences the performance of Sanitizer, we conduct additional experiments and adopt the commonly used Dirichlet distribution~\cite{pmlr-v97-yurochkin19a} to create heterogeneous data partitions across clients with varying degrees of non-IIDness controlled by the concentration parameter $\alpha$. A lower $\alpha$ value indicates higher data skew among clients, and vice versa. We maintain consistent experimental settings and evaluation metrics as detailed in Section~\ref{5-2}. As shown in Table~\ref{tab_non-iid}, Sanitizer maintains robust watermark detection performance across different non-IID conditions, with WDR remaining above 97\%. It is also observed that as the value of $\alpha$ decreases (i.e., data heterogeneity grows), the ASR climbs. This suggests that skewed data distributions may introduce vulnerabilities that slightly compromise the Sanitizer's ability to fully neutralize harmful backdoors, enhancing their effectiveness. Additionally, the increased heterogeneity slightly weakens model utility, a result consistent with previous research findings~\cite{chen2023fedright, FL3}, but overall Sanitizer remains resilient in non-IID settings, prioritizing the reliability of the harmless watermarking functionality.

\subsection{Comparison with Additional Approaches}
\label{Comparison_FedIPR}
To benchmark Sanitizer against more approaches, we next compare it with FedIPR~\cite{FL2}, a state-of-the-art client-side FL watermarking method for ownership verification. FedIPR can also allow each client to embed their own backdoor-based watermark for contribution demonstration, but it may include harmful triggers and lack sanitization to prevent them. To ensure a fair comparison, we solely use backdoor-based watermarks in FedIPR. Besides, we also consider transforming sanitization mechanisms to be applied on the FL-trained model post-FL, without additional mechanisms employed during FL with watermark embedding. This represents an intuitive sanitizing approach that prioritizes efficiency, an efficiency-focused baseline (Efficiency-FB) that incurs no extra overhead during FL. Table~\ref{Comparison} shows the comparison of Sanitizer with FedIPR and Efficiency-FB in a consistent setting. While FedIPR achieves a watermark verification success rate of 99.25\%, this stays the same for both harmful and harmless environments. For Efficiency-FB, while it does not incur extra burden during FL, its success rate for watermark verification is only 21.14\%, whereas that for backdoor attacks remains high at 98.06\%. On the contrary, Sanitizer successfully balances watermark detection, attack resistance, and model performance in our context.

\begin{table}[t]
  \centering
  \normalsize
  \renewcommand{\arraystretch}{1.4}  
  \setlength{\tabcolsep}{9.0pt}  
  \caption{Effectiveness comparison of Sanitizer with two additional methods in consistent settings on CIFAR10 with ResNet18. Besides Sanitizer, neither method is able to satisfy all the properties in our context.}
  \begin{tabular}{cccc}
    \hline
    Approach & \textbf{WDR $\uparrow$} & \textbf{ASR $\downarrow$} & \textbf{ACC $\uparrow$}\\
    \hline
    FedIPR & 99.25\% & 99.25\%  &  86.85\% \\
    Efficiency-FB & 21.14\% & 98.06\% & 83.02\% \\
    \hline
    \rule{0pt}{3ex} Sanitizer & 96.21\% & 11.10\% & 87.28\% \\
    \hline
  \end{tabular}
  \label{Comparison}
\end{table}

\subsection{Parameter Studies}
\label{5-7}
We further investigate the impact of different parameter settings on the performance of Sanitizer, as well as its sensitivity to the variation in them. Our parameter studies mainly examine the effectiveness and efficiency of Sanitizer under two key factors in our setting: (i) varying numbers of reverse engineering epochs per round and (ii) different backdoor subnet sizes. For each factor, we only vary this single factor while keeping all other setups consistent with the default setting of the evaluation in Section~\ref{5-2}. All the parameter experiments in this subsection are performed on CIFAR10 using ResNet18.

\textbf{Impact of Different Round-spread Reverse Engineering Epochs on Efficiency}. 
As shown in Figure~\ref{variant_epochs_all}, we investigate the metrics about efficiency across different round-spread reverse engineering epochs per round: (i) the GPU\_Mem (MB) consumed for processing a single client-submitted model during server-side operations, and (ii) the average Time (s) required for server-side operations per round. We also report the baseline values under the same settings for reference. These values are fixed, as the baseline does not involve the two parameters. According to the results, GPU memory utilization remains consistent, ranging from approximately 190 MB to 194 MB across all epoch settings, which is notably lower than the baseline. This stability suggests that GPU memory usage of Sanitizer exhibits low sensitivity to the number of reverse engineering epochs per round. On the other hand, the time required for server-side operations per round shows a clear upward trend, escalating from 19.44 seconds at 1 epoch to 28.12 seconds at 2 epochs, and further rising to 53.73 seconds at 4 epochs. Such a trajectory in the figures stems from the inherently positive relationship between time and the number of reverse engineering epochs per round. Obviously, it persistently remains below the baseline value under varying epochs. The results show that reducing the number of round-spread reverse engineering epochs can save the server-side operation time, further validating Sanitizer’s optimization for efficiency.

\textbf{Impact of Different Round-spread Reverse Engineering Epochs on Effectiveness}. As illustrated in Figure~\ref{variant_epochs_all}, we investigate the WDR and the ASR under different reverse engineering epochs per communication round. Specifically, both the baseline and Sanitizer consistently maintain a robust and near-perfect WDR over the set of epochs \{1, 2, 4\}. The ASR follows a similar trend, showing a slight and routine difference between the two approaches and remaining stable as the number of epochs varies. Results from this experiment confirm that different round-spread reverse engineering epochs have virtually no impact on the Sanitizer, which remains effective for any value of this factor. This shows that the effectiveness of Sanitizer is not sensitive to the number of round-spread reverse engineering epochs. In light of the aforementioned analysis pertaining to efficiency and effectiveness, it can be concluded that, even though Sanitizer achieves comparable effectiveness to the baseline across varying epochs, the latter is associated with significantly higher costs.

\textbf{Impact of Different Backdoor Subnet Sizes on Efficiency}. Figure~\ref{variant_subnet_all} reports the results. Our primary objective is to investigate the relationship of efficiency between the backdoor subnet size (across three configurations: 10\%, 20\%, and 40\%) and the key efficiency metrics. Sanitizer exhibits an increasing trend in both GPU memory utilization and required process time as the backdoor subnet size grows. In particular, the GPU memory rises from 117 MB to 379 MB, and the process time per round extends from 16.80 seconds to 38.90 seconds. All of these results largely outperform the baseline, demonstrating the intention behind the design of Sanitizer, which aims to enhance efficiency by reducing the backdoor subnet size.

\textbf{Impact of Different Backdoor Subnet Sizes on Effectiveness}. Similarly, we systematically vary the backdoor subnet size and investigate the impact of them on the effectiveness of Sanitizer, keeping other variables constant to ensure fairness in comparisons. Figure~\ref{variant_subnet_all} presents the WDR results as the backdoor subnet size increases. Sanitizer achieves a near-perfect WDR of 98.77\% at the smallest backdoor subnet size (10\%) and maintains a high WDR across all configurations of larger subnet sizes (20\% and 40\%). As for the ASR, Sanitizer shows a much lower ASR than the baseline and a stable trend across three subnet sizes. Hence, while markedly reducing resource and time consumption, Sanitizer achieves comparable performance to the baseline across different backdoor subnet sizes, demonstrating its effectiveness in eliminating harmful backdoor knowledge and enabling contribution demonstration via our harmless watermarks. In summary, the effectiveness of Sanitizer is not sensitive to the reduction in subnet size within a reasonable range brought about by the need to improve efficiency during the FL process.

\section{Conclusions} 
With the goal of strengthening trustworthy FL systems, we have introduced Sanitizer, a novel and efficient sanitization approach designed to address the harmful effects and conflict issues posed by backdoor-based client-side watermarks for contribution demonstration in practical FL applications. Sanitizer leverages backdoor subnets and round-spread reverse engineering to efficiently reverse the backdoor triggers, then confines them to their own harmless, client-specific environment before deployment and eventually makes the watermarks function in harmless ways to successfully verify the contributions. Empirical experiments have demonstrated the effectiveness, harmlessness, and especially the efficiency of Sanitizer. We advocate for further exploration in this direction and believe that Sanitizer will inspire more advanced work for better IP protection in various real-world applications.

\section*{Acknowledgement} 
This research is partially supported by the Croucher Start-up Allowance (Project \#2499102828) and RGC Early Career Scheme (Project \#27211524). Any opinions, findings, or conclusions expressed in this material are those of the authors and do not necessarily reflect the views of the Croucher Foundation and RGC.

\bibliographystyle{IEEEtran}
\bibliography{Refers}

\appendices
\section{Outline of Supplementary Materials}
\noindent This part serves as the appendix, providing additional details for our main paper. It is organized as follows:

\begin{itemize}
    \item \textbf{Section}~\ref{appendix_A0}: Data Availability Statements.
    \item \textbf{Section}~\ref{appendix_A1}: The Algorithm of Sanitizer.
    \item \textbf{Section}~\ref{appendix_A2}: Default Parameter Setting.
    \item \textbf{Section}~\ref{appendix_A3}: More Details on the Extracted Backdoor Subnet.
    \item \textbf{Section}~\ref{appendix_A4}: Visual Examples of Trigger Inversion on the FL-trained Model without Sanitizer.
    \item \textbf{Section}~\ref{appendix_A5}: More Backdoor Trigger Types.
    \item \textbf{Section}~\ref{appendix_A6}: More Analysis and Discussion.
    \item \textbf{Section} \ref{appendix_A7}: Limitation and Future Work.
\end{itemize}

\subsection{Data Availability Statements}
\label{appendix_A0}
We fully support the principles of open science, aiming to promote the transparency, reproducibility, and collaborative research. The datasets (Fashion-MNIST~\cite{Han_Rasul_Vollgraf_2017}, CIFAR10~\cite{krizhevsky2009learning}, CIFAR100~\cite{krizhevsky2009learning}, and TinyImageNet~\cite{le2015tiny}) used in this study are publicly available and widely recognized as the benchmark in the image classification tasks. The details of these datasets are as follows:\\

\begin{algorithm*}[t]
\caption{The Baseline method (Left) \textbf{vs.} Our Sanitizer (Right). The main difference is the \textcolor{darkgreen}{\(\triangleright\) Green Line}.}
\label{alg:algorithm_toxin}
\vspace{-10pt} 
\begin{multicols}{2}
\setlength{\columnsep}{0.3cm} 
\textbf{Input}: Total communication rounds of FL $R$; Number of clients $K$; Defense data $(x_d, y_d) \in \mathcal{D}_d$; Harmless data $x_u \in \mathcal{D}_u$; Poisoning rate of backdoor embedding $\rho\%$;\\
\textbf{Output}: \makebox[0pt][l]{Harmless watermarked FL-trained model $\mathcal{F}(\theta_u)$.}
\begin{algorithmic}[1]
\STATE Server sends $\theta^0_{global}$ to all $K$ clients for initialization.
\FOR{each round $r = 0, 1, \dots, R-1$}
    \STATE \texttt{/* Client-Side */}
    \FOR{each client $k = 0, 1, \dots, K-1$ in parallel}
        \STATE Client $k$ trains a backdoored model $f(\theta_k)$ via Equation~\ref{eq:backdoor_effect} on $\rho\%$.
    \ENDFOR
    \STATE \texttt{/* Server-Side */}
    \FOR{each $f(\theta_k)$, $k = 0, 1, \dots, K-1$ in parallel}
        \STATE Conduct full Reverse Engineering on $f(\theta_k)$ for trigger pattern $\Delta^k_t$ recovery and target class $y^k_t$ detection. \quad \quad \quad \quad \quad \quad \quad \(\triangleright\) Trigger Recovery.
        \STATE Conduct $\min \mathbb{E}_{(\boldsymbol{x}_d, y_d)\in\mathcal{D}_{d}}\mathcal{L}(f\left(x_d + \Delta^k_t;\theta_k\right), y_d)$ on $f(\theta_k)$.  \quad \quad \quad \quad \quad \quad \quad \(\triangleright\) Unlearning.
    \ENDFOR
    \STATE Aggregation to update $\theta^r_{global}$ for the $r+1$ round.
\ENDFOR

\STATE Server conducts $\min \mathbb{E}_{x_u\in\mathcal{D}_{u}}\mathcal{L}(f\left(x_u + \Delta;\theta\right), y_t)$ on FL-trained Model $\mathcal{F}(\theta)$. \quad \quad \quad \quad \quad \quad \(\triangleright\) Relearning.
\RETURN Harmless watermarked FL-trained model $\mathcal{F}(\theta_u)$.
\end{algorithmic}

\columnbreak

\textbf{Input}: Total communication rounds of FL $R$; Number of clients $K$; Defense data $(x_d, y_d) \in \mathcal{D}_d$; Harmless data $x_u \in \mathcal{D}_u$; Rate of backdoor subnet size $s\%$; Epoch of round-spread reverse engineering per round $e$; Poisoning rate of backdoor embedding $\rho\%$;\\
\textbf{Output}: \makebox[0pt][l]{Harmless watermarked FL-trained model $\mathcal{F}(\theta_u)$.}
\begin{algorithmic}[1]
\STATE Server sends $\theta^0_{global}$ to all $K$ clients for initialization.
\FOR{each round $r = 0, 1, \dots, R-1$}
    \STATE \texttt{/* Client-Side */}
    \FOR{each client $k = 0, 1, \dots, K-1$ in parallel}
        \STATE Client $k$ trains a backdoored model $f(\theta_k)$ via Equation~\ref{eq:backdoor_effect} on $\rho\%$.
    \ENDFOR
    \STATE \texttt{/* Server-Side */}
    \FOR{each $f(\theta_k)$, $k = 0, 1, \dots, K-1$ in parallel}
        \STATE \textcolor{darkgreen}{\(\triangleright\)} Conduct Clean Unlearning on $f(\theta_k)$ with $(x_d, y_d)$ by Equation~\ref{eq:unlearning_substep} to obtain $f(\theta'_k)$.
        \STATE \textcolor{darkgreen}{\(\triangleright\)} Identify and extract a backdoor subnet $f_s(\theta_k^*)$ from $f(\theta_k)$ of Top $s\%$ UWC via Equation~\ref{eq:uwc_substep}.
        \STATE \textcolor{darkgreen}{\(\triangleright\)} Conduct round-spread reverse engineering for $e$ epochs on $f_s(\theta_k^*)$ for trigger pattern $\Delta_r$ and target class $y_r$ at $r$ round and pruning.
    \ENDFOR
    \STATE Aggregation by Equation~\ref{eq:fedavg_substep} for the $r+1$ round.
\ENDFOR
\STATE Server conducts Harmless Relearning on FL-trained Model $\mathcal{F}(\theta)$ via Equation~\ref{eq:Relearning_substep2}.
\RETURN Harmless watermarked FL-trained model $\mathcal{F}(\theta_u)$.
\end{algorithmic}
\end{multicols}
\end{algorithm*}

\begin{itemize}
    \item \textbf{Fashion-MNIST}: Fashion-MNIST contains 70,000 28$\times$28 grayscale images across 10 fashion categories. The dataset is split into 60,000 training images and 10,000 test images. It is designed as a modern alternative to the classic MNIST dataset. It can be accessed from \url{https://github.com/zalandoresearch/fashion-mnist}. It is supported natively by PyTorch and will be downloaded automatically.\\
    
    \item \textbf{CIFAR10}: CIFAR10 consists of 60,000 32$\times$32 color images in 10 different classes. The dataset is divided into 50,000 training images and 10,000 test images, and is widely used for evaluating machine learning models on small-scale object recognition tasks. CIFAR10 is publicly available at \url{http://www.cs.toronto.edu/~kriz/cifar.html}, which is supported natively by PyTorch and will be downloaded automatically.\\
    
    \item \textbf{CIFAR100}: CIFAR100 consists of 60,000 32$\times$32 color images categorized into 100 different classes. The dataset is split into 50,000 training images and 10,000 test images, and, like CIFAR10, it is widely used for benchmarking machine learning models on small-scale object recognition tasks. It is publicly available at \url{http://www.cs.toronto.edu/~kriz/cifar.html}, and is natively supported by popular frameworks such as PyTorch, which can automatically download the dataset.\\
    
    \item \textbf{TinyImageNet}: TinyImageNet includes 100,000 64$\times$64 color images across 200 classes.  Each class contains 500 training images and 50 validation images. TinyImageNet is a smaller, more manageable subset of the full ImageNet dataset, commonly used for evaluating deep learning models. It can be downloaded from \url{http://cs231n.stanford.edu/tiny-imagenet-200.zip}.
\end{itemize}

\begin{table}[t]
    \centering
    \large  
    \renewcommand{\arraystretch}{1.15} 
    \resizebox{0.92\columnwidth}{!}{ 
    \begin{tabular}{|c|c|c|}
        \hline
        \textbf{Module} & \textbf{Parameter} & \textbf{Setting} \\
        \hline
          & communication rounds & 200 \\
          & local training epochs & 10 \\
          & local training batch size & 128 \\
        Client-side  & local learning rate & 0.05 \\
        Training  & optimizer & SGD \\
        Process  & momentum & 0.9 \\
          & weight decay & 5e-4 \\
          & backdoor poisoning rate & 0.1 \\
        \hline
          & defense data rate & 0.05 \\
          & clean unlearning epochs & 10 \\
        Clean  & clean unlearning batch size & 128 \\
        Unlearning & learning rate & 0.01 \\
          & early stop threshold & 0.15 \\
        \hline
          & backdoor subnet size rate & 0.2 \\
        Round-spread & round-spread epochs & 2 \\
        Reverse & batch size & 128 \\
        Engineering & learning rate & 0.2  \\
          & weight for L1 norm of mask & 0.01 \\
        \hline
          & relearning epochs & 50 \\
          & batch size & 64 \\
        Harmless & learning rate & 0.005 \\
        Relearning & weight for clean objective  & 0.8 \\
          & weight for harmless objective & 0.2 \\
        \hline
    \end{tabular}}
    \caption{Default Parameter Setting in FL with Sanitizer.}
    \label{table:fl_basic_parameters}
\end{table}

\subsection{The Algorithm of Sanitizer}
\label{appendix_A1}
In Algorithm~\ref{alg:algorithm_toxin}, we present the algorithmic details that we introduce in Section~\ref{sec: Methodology}. We describe the baseline method as well as Sanitizer in full. The baseline employs a naive sanitization method, which, while effective to some extent, suffers from unacceptable inefficiency. This limitation motivated the development of our proposed Sanitizer. By maintaining the sanitization effectiveness, Sanitizer significantly reduces unnecessary overhead, leading to noticeable improvements in both processing time and GPU memory consumption. These optimizations make it a far more practical solution in scenarios requiring both efficiency and effectiveness.

\subsection{Default Parameter Setting}
\label{appendix_A2}
Table~\ref{table:fl_basic_parameters} summarizes the default parameter setting in our empirical evaluation.

\subsection{More Details on the Extracted Backdoor Subnet}
\label{appendix_A3}
Based on the default settings, we present a further analysis of the extracted backdoor subnet in two different architectures when applied to CIFAR10, which has a resolution of 32$\times$32 pixels and three color channels (RGB). As shown in Table~\ref{table_params_appendix}, we compared the number of parameters and FLOPs between the original network and the extracted backdoor subnet. The results clearly demonstrate a significant reduction in computational complexity. This decrease in model size and operations highlights the efficiency gained through subnet extraction, making it more suitable for resource-constrained environments without sacrificing key performance metrics.

\subsection{Visual Examples of Trigger Inversion on the FL-trained Model without Sanitizer}
\label{appendix_A4}
We observe that recovering the trigger from an FL-trained model without employing Sanitizer appears to be not feasible. In this scenario, no defense mechanisms are applied during the training phase of the existing backdoor-based client-side watermarked FL, and reverse engineering is only performed on the final FL-trained model after the FL process has completed, prior to deployment. Table~\ref{table_visual_appendix} illustrates that the shortcut, or trigger, associated with each class cannot be successfully recovered in the final FL-trained model. We conjecture that this outcome may be attributed to the aggregation process, which potentially dilutes the backdoor effect, thereby obscuring the true trigger. This dilution likely results from the averaging of model updates across clients, which disperses the influence of individual backdoor triggers and diminishes their impact on the global model. Consequently, the aggregated model fails to retain the malicious characteristics necessary for trigger identification. The results indicate that it seems not possible to recover the specific triggers associated with each target label, highlighting the difficulty of accurately identifying and extracting the embedded backdoors under this condition.

\begin{table}[h] 
  \centering
  \renewcommand{\arraystretch}{1.4}
  \small
  \caption{Complexity of the extracted backdoor subnet in three different architectures when applied to CIFAR10.}
  \begin{subtable}[t]{0.48\textwidth}
      \centering
      \begin{tabular}{@{}>{\centering\arraybackslash}p{2.3cm}@{}>{\centering\arraybackslash}p{2.7cm}@{}>{\centering\arraybackslash}p{2.7cm}@{}}
      \toprule
      \multirow{2}{*}{\textbf{Network}} & \multicolumn{2}{c}{\textbf{CIFAR10 \& ResNet18}} \\
      \cmidrule(lr){2-3}
       & \textbf{\#Params} & \textbf{FLOPs} \\ 
      \midrule
      \footnotesize \shortstack{Original Network} & \normalsize \textcolor{red}{11.17M} & \normalsize \textcolor{red}{1.116G} \\ 
      \cmidrule(lr){1-3}
      \footnotesize \shortstack{Backdoor Subnet} & \normalsize \textcolor{darkgreen}{444.22K} & \normalsize \textcolor{darkgreen}{43.677M} \\
      \bottomrule
      \end{tabular}
      \label{table_cifar10_comparison_1}
  \end{subtable}%

  \vspace{15pt}

  \begin{subtable}[t]{0.48\textwidth}
      \centering
      \begin{tabular}{@{}>{\centering\arraybackslash}p{2.3cm}@{}>{\centering\arraybackslash}p{2.7cm}@{}>{\centering\arraybackslash}p{2.7cm}@{}}
      \toprule
      \multirow{2}{*}{\textbf{Network}} & \multicolumn{2}{c}{\textbf{CIFAR10 \& MobileNetV3}} \\
      \cmidrule(lr){2-3}
      & \textbf{\#Params} & \textbf{FLOPs}\\ 
      \midrule
      \footnotesize \shortstack{Original Network} & \normalsize \textcolor{red}{1.684M} & \normalsize \textcolor{red}{5.145M} \\ 
      \cmidrule(lr){1-3}
      \footnotesize \shortstack{Backdoor Subnet} & \normalsize \textcolor{darkgreen}{80.821K} & \normalsize \textcolor{darkgreen}{336.630K}\\
      \bottomrule
      \end{tabular}
      \label{table_tiny_comparison_1}
  \end{subtable}
  
  \vspace{15pt}

  \begin{subtable}[t]{0.48\textwidth}
      \centering
      \begin{tabular}{@{}>{\centering\arraybackslash}p{2.3cm}@{}>{\centering\arraybackslash}p{2.7cm}@{}>{\centering\arraybackslash}p{2.7cm}@{}}
      \toprule
      \multirow{2}{*}{\textbf{Network}} & \multicolumn{2}{c}{\textbf{CIFAR10 \& ViT-Base/2-32}} \\
      \cmidrule(lr){2-3}
       & \textbf{\#Params} & \textbf{FLOPs} \\ 
      \midrule
      \footnotesize \shortstack{Original Network} & \normalsize \textcolor{red}{16.552M} & \normalsize \textcolor{red}{8.535G} \\ 
      \cmidrule(lr){1-3}
      \footnotesize \shortstack{Backdoor Subnet} & \normalsize \textcolor{darkgreen}{639.547K} & \normalsize \textcolor{darkgreen}{332.993M} \\
      \bottomrule
      \end{tabular}
      \label{table_cifar10_comparison_2}
  \end{subtable}
  
  \label{table_params_appendix}
\end{table}

\begin{table}[h]
\centering
\normalsize
\caption{The Original Trigger corresponding to each class and its recovered version.}
\begin{tabular}{@{}>{\centering\arraybackslash}m{1.6cm}@{}>{\centering\arraybackslash}m{2.1cm}@{}>{\centering\arraybackslash}m{2.1cm}@{}>{\centering\arraybackslash}m{2.1cm}@{}}
\toprule
\normalsize  & \normalsize Class 1 & \normalsize Class 2 & \normalsize Class 3 \\
\midrule
Original Trigger & \includegraphics[width=1.9cm, height=1.9cm]{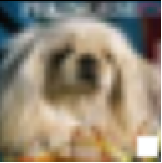} & \includegraphics[width=1.9cm, height=1.9cm]{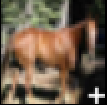} & \includegraphics[width=1.9cm, height=1.9cm]{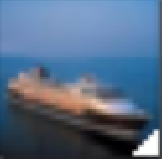} \\
\cmidrule(lr){1-4}
Reversed Version & \includegraphics[width=1.9cm, height=1.9cm]{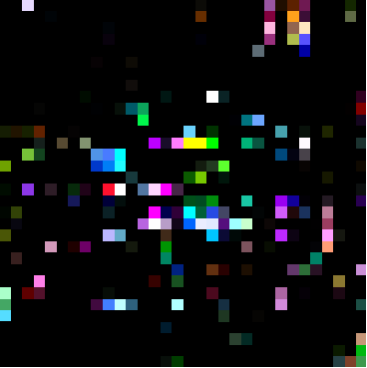} & \includegraphics[width=1.9cm, height=1.9cm]{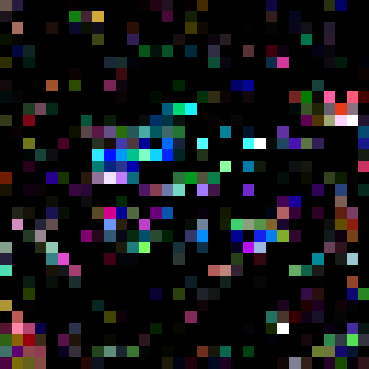} & \includegraphics[width=1.9cm, height=1.9cm]{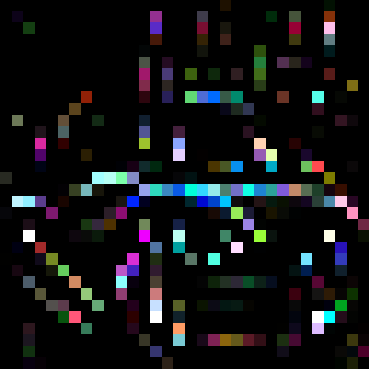} \\
\bottomrule
\end{tabular}
\label{table_visual_appendix}
\end{table}

\begin{table}[t]
  \centering
  \normalsize
  \renewcommand{\arraystretch}{1.4}  
  \setlength{\tabcolsep}{9.0pt}  
  \caption{Experiments with other trigger types in default settings, Blend and TrojanNN. Sanitizer also demonstrates considerable effectiveness.}
  \begin{tabular}{cccc}
    \hline
    Approach & \textbf{WDR $\uparrow$} & \textbf{ASR $\downarrow$} & \textbf{ACC $\uparrow$}\\
    \hline
    Blend & 95.92\% & 13.23\%  &  85.46\% \\
    TrojanNN & 98.63\% & 16.42\% & 82.62\% \\
    \hline
  \end{tabular}
  \label{Trigger-Types}
\end{table}

\subsection{More Backdoor Trigger Types}
\label{appendix_A5}
Apart from Badnet, we conduct additional experiments to evaluate how Sanitizer performs when clients adopt different data poisoning backdoor techniques, particularly with respect to trigger types. We analyze: (i) Blend~\cite{chen2017targetedblend}, which superimposes a partially transparent image as a trigger, and (ii) TrojanNN~\cite{liu2018trojaningtrojannn}, which formulates an optimization problem to identify an ideal trigger pattern. Sanitizer also demonstrates considerable effectiveness. As shown in Table~\ref{Trigger-Types}, like Badnet, Sanitizer enables watermark verification in designated, harmless environments with a success rate exceeding 95.92\%, , while the same triggers can only manipulate model decisions (in a harmful manner) with a success rate of less than 16.42\%. The ACCs are maintained at an appreciable level.

\subsection{More Analysis and Discussion}
\label{appendix_A6}

\noindent $\bullet$ The number of communication rounds required to achieve both optimal ACC for the main task and successful backdoor trigger inversion is also a factor influencing the overall efficiency of FL. Reducing the round-spread reverse engineering epochs may lead to an increased number of communication rounds required to achieve the desired outcomes, and vice versa. This highlights a clear trade-off between minimizing communication rounds and optimizing server-side efficiency. This trade-off requires careful consideration to balance computational resources and performance goals effectively.\\

\noindent $\bullet$ Why mainly assume a Badnet-style backdoor? We argue that in a scenario where FL clients co-own the jointly trained model, clients tend to embed watermarks in the simplest and most direct manner to demonstrate their contribution, obviating the need for complex and impractical techniques. As clients cannot easily break into the source code and manipulate the standard local learning, the easiest way for them to watermark the model in FL is to conduct data poisoning backdoors because it avoids tampering with the optimization process, which many sophisticated techniques proposed for centralized learning require. This means they don't need to expend significant effort (or resources) to demonstrate their contribution; instead, a more practical and straightforward approach for them is implementing a Badnet-like patch replacement that applies a trigger pattern onto an image and changes its label. It is reasonable to assume that they prefer to demonstrate their contribution with minimal cost rather than utilizing more sophisticated backdoor embedding methods. Moreover, we specifically focus on Badnet-like backdoor-based watermarks, as our approach is not designed to defend against backdoors in general, but rather to serve as a safeguard when such backdoors are used as watermarks under specific conditions.

\subsection{Limitation and Future Work}
\label{appendix_A7}

Sanitizer has confirmed the promising results of making the backdoor-based client-side watermarks harmless to enhance the contribution demonstration in FL, but it also comes with several limitations: (a) Sanitizer has not yet optimized the computational cost associated with the subsequent step after FL, although it is relatively negligible compared to the whole FL period; (b) Sanitizer has not yet been applied to tasks beyond image classification.

In the future, we will address the challenges and further extend the applicability of Sanitizer. Sanitizer represents the first research effort that directly targets the harmlessness of backdoor-based client-side watermarks in FL. This work not only advances the state-of-the-art for image classification tasks but also paves the way for future exploration across a broader range of ML applications, including object detection and beyond. We anticipate that the principles behind Sanitizer can be effectively extended to other FL scenarios, like personalized FL, driving further innovations in enhancing the IP protection and contribution demonstration of the FL-trained model.

\end{document}